\documentclass[12pt]{article}

\usepackage{graphicx,amsmath,amssymb}
\usepackage{color,scicite}
\usepackage{chemfig,chemmacros}
\usepackage{url}
\usepackage{gensymb}
\usepackage{pdfpages}
\usepackage{stackengine}

\newenvironment{sciabstract}{%
\begin{quote} \bf}
{\end{quote}}

\newcounter{lastnote}
\newenvironment{scilastnote}{%
\setcounter{lastnote}{\value{enumiv}}%
\addtocounter{lastnote}{+1}%
\begin{list}%
{\arabic{lastnote}.}
{\setlength{\leftmargin}{.22in}}
{\setlength{\labelsep}{.5em}}}
{\end{list}}

\usepackage{newfloat}
\DeclareFloatingEnvironment{efigure}
\DeclareFloatingEnvironment{etable}

\usepackage{times}

\newcommand{\kepler}{{\it Kepler}}

\title{The Origin of RNA Precursors on Exoplanets}

\author{Paul B. Rimmer$^{1,2\ast}$ \and Jianfeng Xu$^{2}$ \and Samantha J. Thompson$^{1}$ \and Ed Gillen$^{1}$ \and John D. Sutherland$^{2}$ \and Didier Queloz$^{1}$\\
\normalsize{$^{1}$Cavendish Astrophysics, University of Cambridge,}\\
\normalsize{JJ Thomson Avenue, Cambridge CB3 0HE, UK}\\
\normalsize{$^{2}$MRC Laboratory of Molecular Biology,}\\
\normalsize{Francis Crick Ave, Cambridge CB2 0QH, UK}\\
\normalsize{$^\ast$To whom correspondence should be addressed; E-mail: pbr27@cam.ac.uk.}
}

\begin{document}

\baselineskip24pt

\maketitle

\begin{sciabstract}
Given that the macromolecular building blocks of life were likely produced photochemically in the presence of ultraviolet (UV) light, we identify some general constraints on which stars produce sufficient UV for this photochemistry. We estimate how much light is needed for the UV photochemistry by experimentally measuring the rate constant for the UV chemistry (`light chemistry', needed for prebiotic synthesis) versus the rate constants for the bimolecular reactions that happen in the absence of the UV light (`dark chemistry'). We make these measurements for representative photochemical reactions involving SO$_3^{2-}$ and HS$^-$. By balancing the rates for the light and dark chemistry, we delineate the ``abiogenesis zones'' around stars of different stellar types based on whether their UV fluxes are sufficient for building up this macromolecular prebiotic inventory. We find that the SO$_3^{2-}$ `light chemistry' is rapid enough to build up the prebiotic inventory for stars hotter than K5 (4400 K). We show how the abiogenesis zone overlaps with the liquid water habitable zone. Stars cooler than K5 may also drive the formation of these building blocks if they are very active. The HS$^-$ `light chemistry' is too slow to work even for the Early Earth.
\end{sciabstract}

\section*{Introduction}
Dozens of exoplanets have been found within the liquid water habitable zones of their host stars \cite{Kane2016,Anglada2016,Gillon2017,Dittmann2017}. Living organisms could potentially thrive on the subset of these planets with stable atmospheres, but could life start on these planets in the first place? This question can in principle be answered in terms of whether the conditions necessary for life under a given scenario could be plausibly met on another planet, given what we know about that planet. We can delineate an ‘abiogenesis zone’, outside of which life is unlikely to have originated in the way the given scenario describes. We will focus on the photochemical scenarios presented by Patel et al. (2015,\cite{Patel2015}) and Xu et al. (2018,\cite{Xu2018}), which we find compelling because these are currently the only known prebiotically plausible chemical networks to selectively achieve high yields of nucleosides, amino acids and lipid precursors from chemical initial conditions that are realistic for the Early Earth. Although this network details how the building blocks of life may have been formed, and not the method of their assembly, this first step is a necessary condition for life's origin \cite{Sutherland2017}. 

It is fortuitous that the scenario we choose to investigate is connected to the light of the host star, the one thing we know best about any exoplanet system. We can connect the prebiotic chemistry to the stellar ultraviolet (UV) spectrum to determine whether these reactions can happen on rocky planets around other stars. Ranjan \& Sasselov \cite{Ranjan2016} have begun to explore whether life could arise photochemically on rocky planets around other stars, within the context of this scenario, and show that it is important to know the wavelength range over which this chemistry takes place. Experiments have shown that this critical wavelength range for these reactions is between 200 - 280 nm \cite{Todd2018}.  A comparatively small amount of the ultraviolet light in this range is produced by ultracool stars, raising the question of whether the prebiotic inventory could ever arise on planets around these stars without the help of flares \cite{Rugheimer2015b,Ranjan2017c}. 

\begin{figure}
\includegraphics[trim={0 6cm 6cm 0},clip,width=\linewidth]{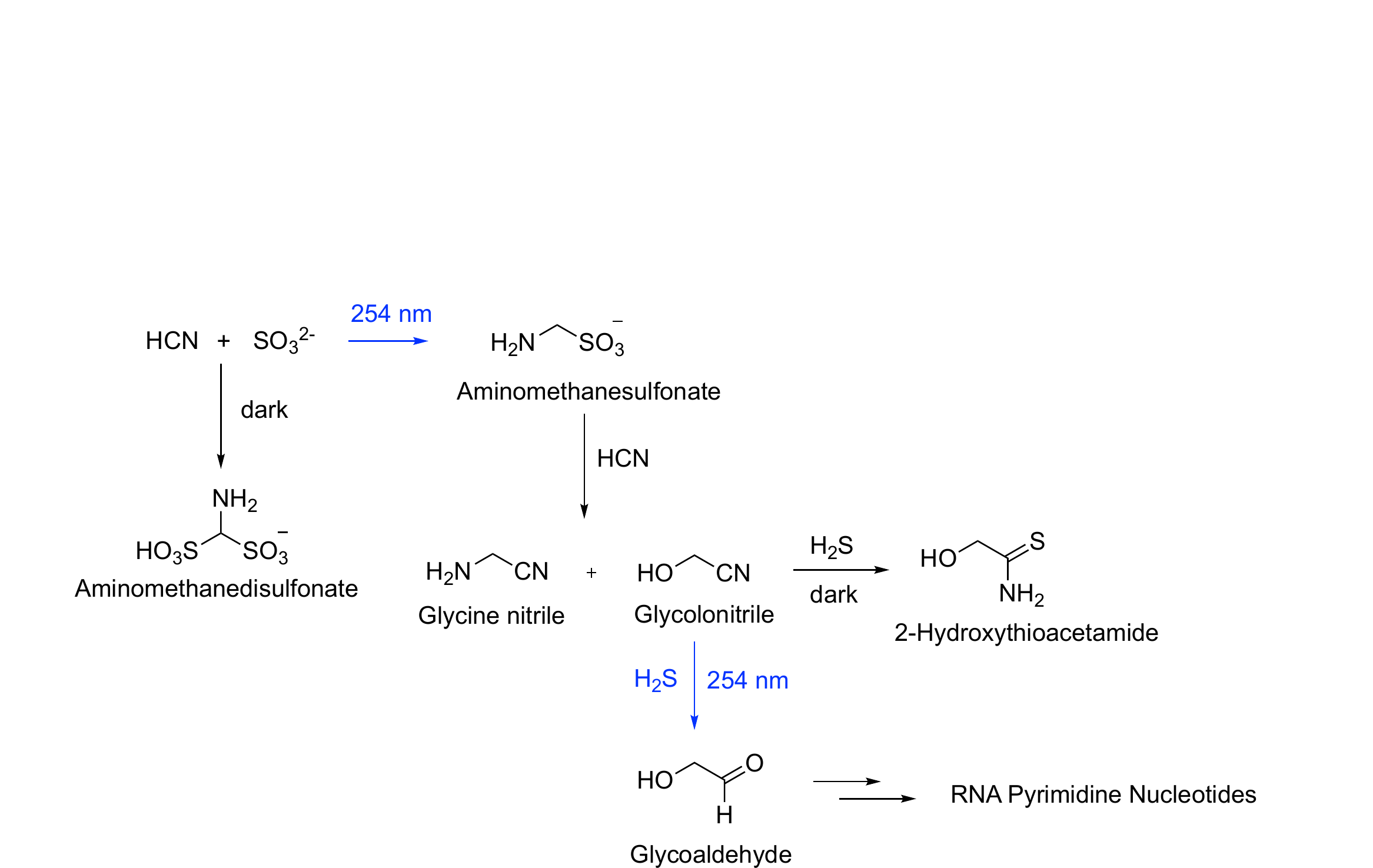}
\caption{{\bf Reaction Scheme.} The scheme of light and dark reactions we explore, considering two steps along the path to forming pyrimidines (RNA precursors), one with bisulfite (SO$_3^{2-}$) as the electron donor, and the other with hydrogen sulfide (H$_2$S, HS$^-$) as the electron donor.}
\label{fig:light-vs-dark}
\end{figure}

To assess where the Patel \cite{Patel2015} and Xu \cite{Xu2018} chemistry can realistically occur, we focus on key reactions along the seven step pathway to form the pyrimidine nucleotide RNA precursors.' This formation is driven by UV detachment of electrons from anions in solution, such as HS$^-$ from H$_2$S \cite{Patel2015} and SO$_3^{2-}$ from SO$_2$ \cite{Xu2018}, in the presence of HCN. The HS$^-$ reaction is representative of the Patel et al. (2015) network and the SO$_3^{2-}$ reaction is representative of the Xu et al. (2018) network, in the sense that each of the photochemical reactions take approximately the same amount of time under the exposure of the UV lamp used in both studies (\cite{Patel2015,Xu2018}, their Supplementary Material). Fig. \ref{fig:light-vs-dark} shows the photochemical products, as well as inert adducts that build up when the light is absent. These inert adducts are sometimes useful for synthesizing amino acids, but cannot lead to the formation of pyrimidine nucleotides. The rate constants for the photochemical reactions (`light chemistry') and the rate constants for the bimolecular reactions that take place in the absence of the UV lamp (`dark chemistry') can be used to estimate the yield of the `light chemistry' products. We define the abiogenesis zone as the zone in which a yield of 50\% for the photochemical products is obtained, adopting the current UV activity as representative of the UV activity during the stellar life time and assuming a young Earth atmosphere.

The choice of a 50\% yield is mandated by the `arithmetic demon', a well-known problem in synthetic chemistry involving many steps (e.g., footnote 8 of \cite{Crispino1993}). Since in this scenario pyrimidine RNA precursors are needed building blocks for life, and since the pathway to form these pyrimidines involves seven steps with similar photochemical rate constants with three of these steps require two of these photochemical products reacting together, the final yield of the pyrimidines from a starting mixture of HCN and either SO$_3^{2-}$ or HS$^-$ will be the yields of each prior step multiplied together. A 10\% yield at each stage will result in an overall pyrimidine yield of $10^{-8}$\%; a 30\% yield for each step in a $6 \times 10^{-4}$\% overall yield, and a 50\% yield for each step in a 0.1\% overall yield. Only the 50\% yield per step is high enough for a robust prebiotic chemistry.

\section*{Results}
We now show our results from comparing the ``light chemistry'' and ``dark chemistry'', in order to identify which environments the photochemical synthesis of pyrimidines is possible, for both SO$_3^{2-}$ and HS$^-$. We found that the SO$_3^{2-}$ dark chemistry was somewhat slower than the HS$^-$ dark chemistry at low temperatures. Additionally, the light chemistry is far more rapid for SO$_3^{2-}$. The rate constants for the dark reactions, resulting in the adducts shown in Fig. \ref{fig:light-vs-dark}, are:
\begin{align}
k_{\rm add}(\text{SO$_3^{2-}$ + HCN}) &= (1.3 \pm 0.5) \times 10^8 \text{ mol$^{-1}$ L s$^{-1}$} \; e^{-(9500 \pm 106) \, \mathrm{K}/T} ,\\
k_{\rm add}(\text{HS$^{-}$ + Glycolonitrile}) &= (2.0 \pm 0.7) \times 10^5 \text{ mol$^{-1}$ L s$^{-1}$} \; e^{-(7500 \pm 130) \, \mathrm{K}/T} ,
\label{eqn:rate-coefficients}
\end{align}

Under the assumption that the cross-sections are constant between 200 and 280 nm (similar to Todd et al. 2018,\cite{Todd2018}), we find cross-sections for SO$_3^{2-}$ and HS$^-$ to produce their respective photoproducts shown in Fig. \ref{fig:light-vs-dark} are:
\begin{align}
\sigma_{\nu}(\text{SO$_3^{2-}$ + HCN}) &= (1.5 \pm 0.3) \times 10^{-21} \, \text{cm$^2$}, \\
\sigma_{\nu}(\text{HS$^{-}$ + Glycolonitrile}) &= (2.0 \pm 0.4) \times 10^{-23} \, \text{cm$^2$}.
\end{align}

\begin{figure}
\includegraphics[width=\linewidth]{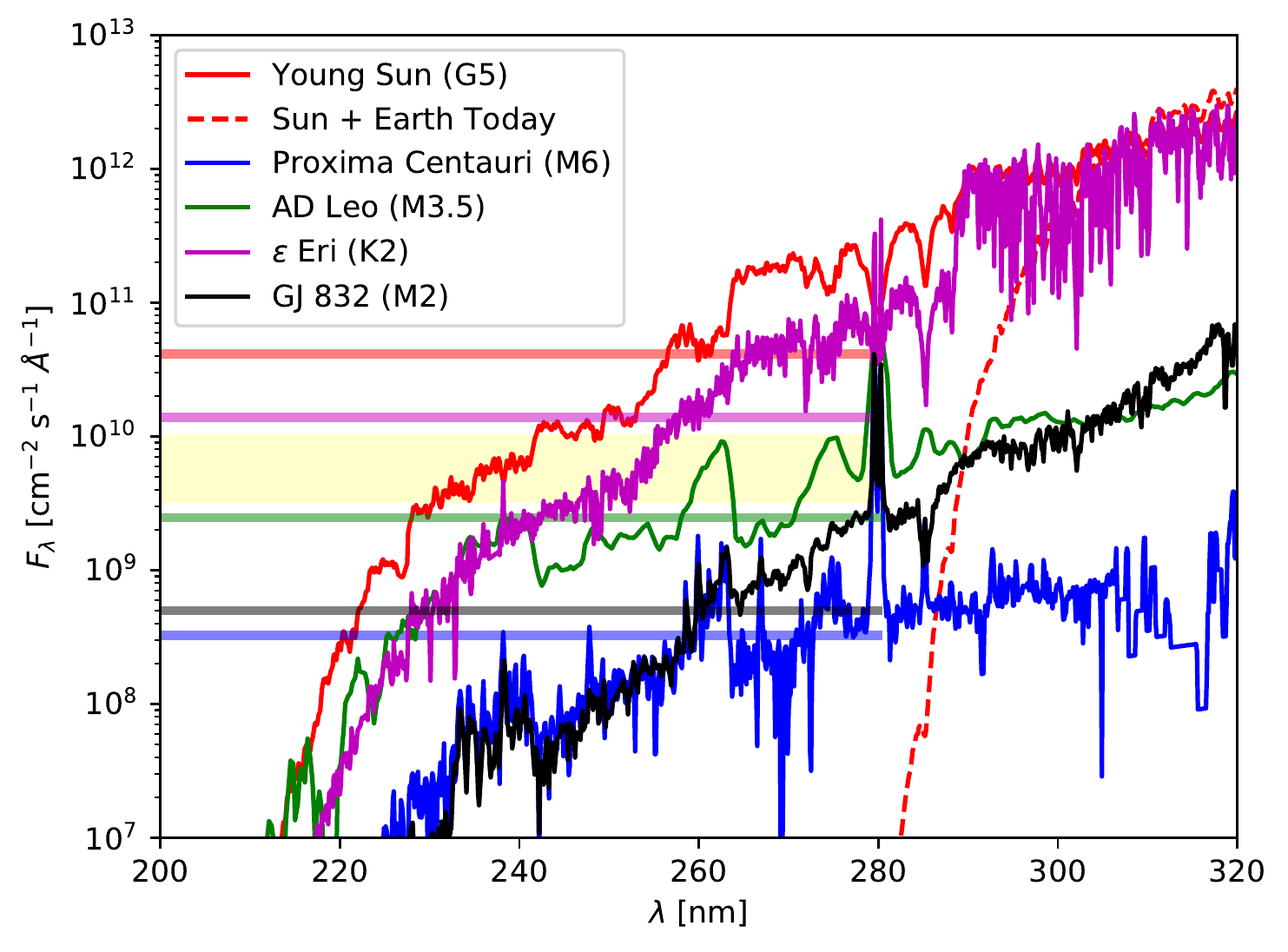}
\caption{{\bf Surface Actinic Fluxes.} The surface actinic flux $F_{\lambda}$ as a function of the wavelength $\lambda$ for planets within the habitable zones of five stars (with spectral type given in the legend): the Early Earth in the case of the young Sun and Proxima b in the case of Proxima Cen. For GJ832, $\epsilon$ Eri and AD Leo, these are hypothetical planets at the innermost edge of the liquid water habitable zone, in order to maximize the surface flux. The yellow shaded region (the width of which accounts for the errors, see Materials and Methods, esp. Eq. (27)-(30)) shows the average flux needed between 200 and 280 nm for the light chemistry and dark chemistry to proceed both at the same rate at 0$^{\circ}$C. The color-shaded horizontal lines show the average fluxes of the respective stars. The surface actinic flux of the Earth today is also included. Its 200-280 nm flux is strongly attenuated by atmospheric ozone.}
\label{fig:UVcomp-blocked}
\end{figure}

\begin{figure}
\centering
\includegraphics[width=0.75\linewidth]{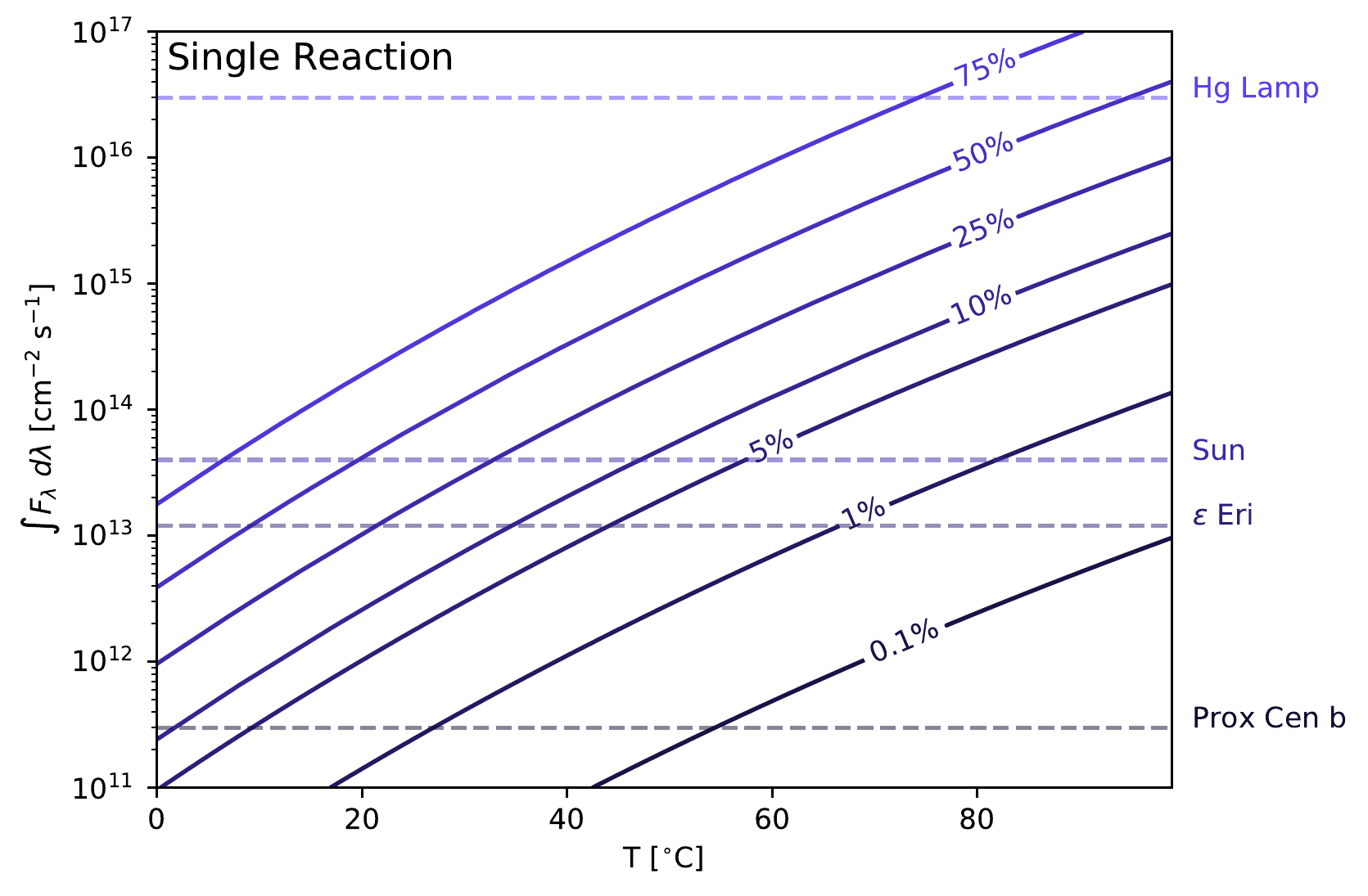}\\
\includegraphics[width=0.75\linewidth]{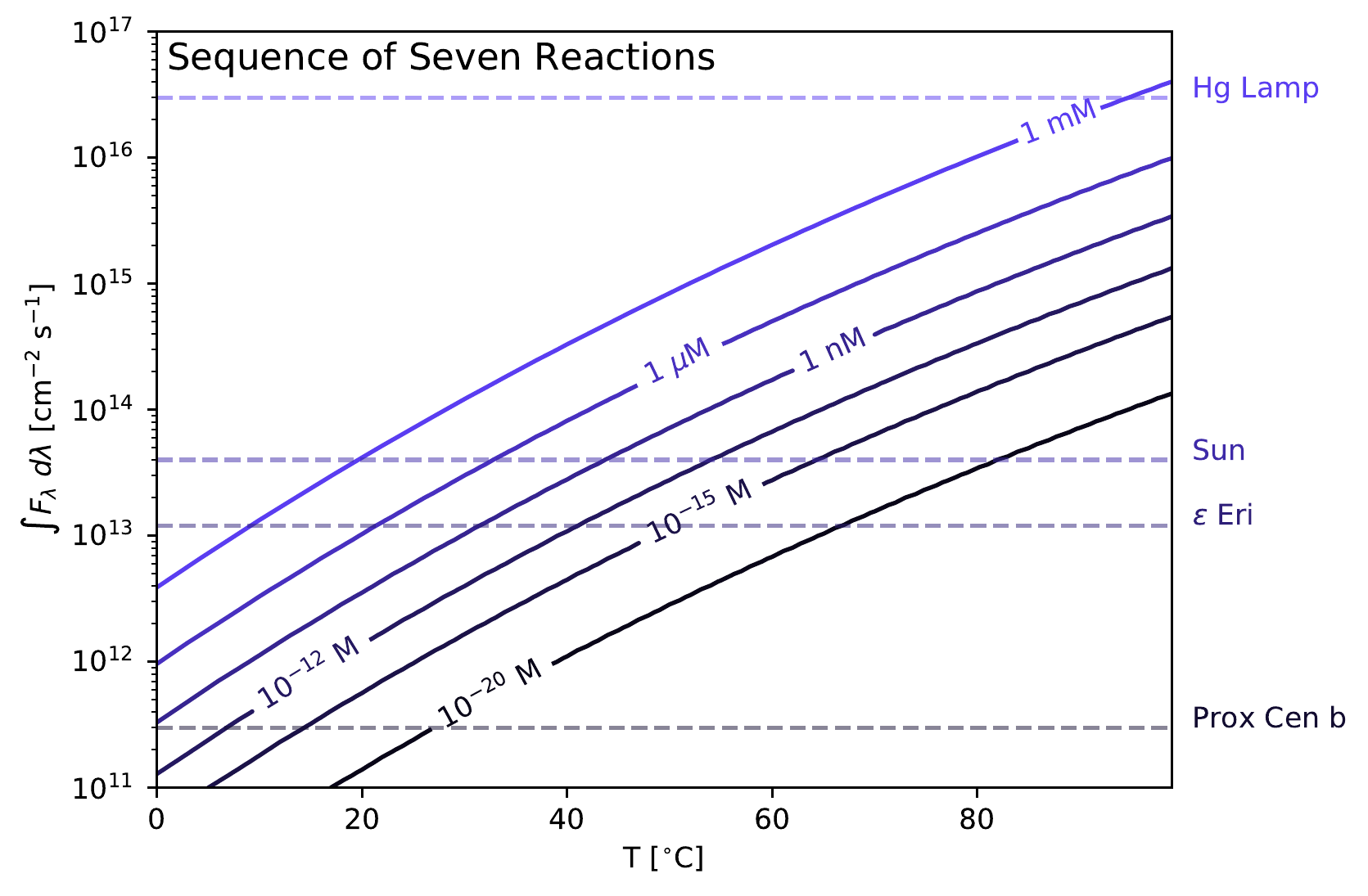}
\caption{{\bf Reaction Yields.} Countour plot of (a) the yield after one reaction and (b) the final concentration of the pyrimidines after seven consecutive reactions (three of which involve two products of a prior reaction) with initial reactants all at 1 M concentrations, as a function of temperature $T$ [K] and the integrated surface UV flux $\int F_{\lambda} \; d\lambda$ [cm$^{2}$ s$^{-1}$] from 200-280 nm, assuming all photochemical reactions have similar photochemical cross-sections and rate constants to the SO$_3^{2-}$ reactions we measured, by multiplying the yield with itself ten times, which accounts for the `arithmetic demon' for the case of ten consecutive reactions, and matches with the pseudo-equilibrium results if rate equations were applied. Comparable surface fluxes are listed on the left.}
\label{fig:yield}
\end{figure}

The dark chemistry is much slower at low temperatures. Since liquid water is a requirement for both the light and dark chemistry to proceed, we consider the best case for the light chemistry at a surface temperature of 0 \degree C. In the case of SO$_3^{2-}$, the light chemistry and dark chemistry move at the same speed when the average actinic flux at the planet's surface between 200 nm and 280 nm is $\sim 6 \times 10^9$ cm$^{-2}$ s$^{-1}$ \AA$^{-1}$. Fig. \ref{fig:UVcomp-blocked} shows surface actinic fluxes of planets with 80\% N$_2$, 20\% CO$_2$, 0.1\% H$_2$O atmospheres with surface pressure 1 bar (a plausible abiotic atmosphere similar to what was likely the atmosphere of the Earth 3.5 Gya \cite{Kasting1993}), within their liquid water habitable zones of a variety of stars of a variety of spectral types, compared to this limiting flux. The modern-day Earth with its 80\% N$_2$, 20\% O$_2$ atmosphere is also included in this figure. Fig. \ref{fig:yield} shows the yield from a single reaction, and the final concentration of pyrimidine concentration from a sequence of ten reactions, driven by SO$_3^{2-}$ photodetachment, as a function of surface actinic flux and temperature.

This comparison shows that the yield for the light chemistry, for each step, exceeds 50\%, and therefore the Early Earth meets our criterion for lying within the abiogenesis zone. The SO$_3^{2-}$ dark chemistry is slow enough that the prebiotic inventory can be built up in regions where the surface temperature is less than $\sim 20$ \degree C. The HS$^-$ dark chemistry is more rapid and, coupled with the much smaller photodetachment cross-section, the 50\% yield is not achieved for the HS$^-$ reaction even at 0 \degree C on the Early Earth. The photodetachment of SO$_3^{2-}$ is sufficiently rapid that the prebiotic inventory can be photochemically generated in a 0 \degree C environment from the quiescent flux of a K5 dwarf ($T_{\rm eff} \approx 4400$ K).

\begin{figure}
\includegraphics[width=\linewidth]{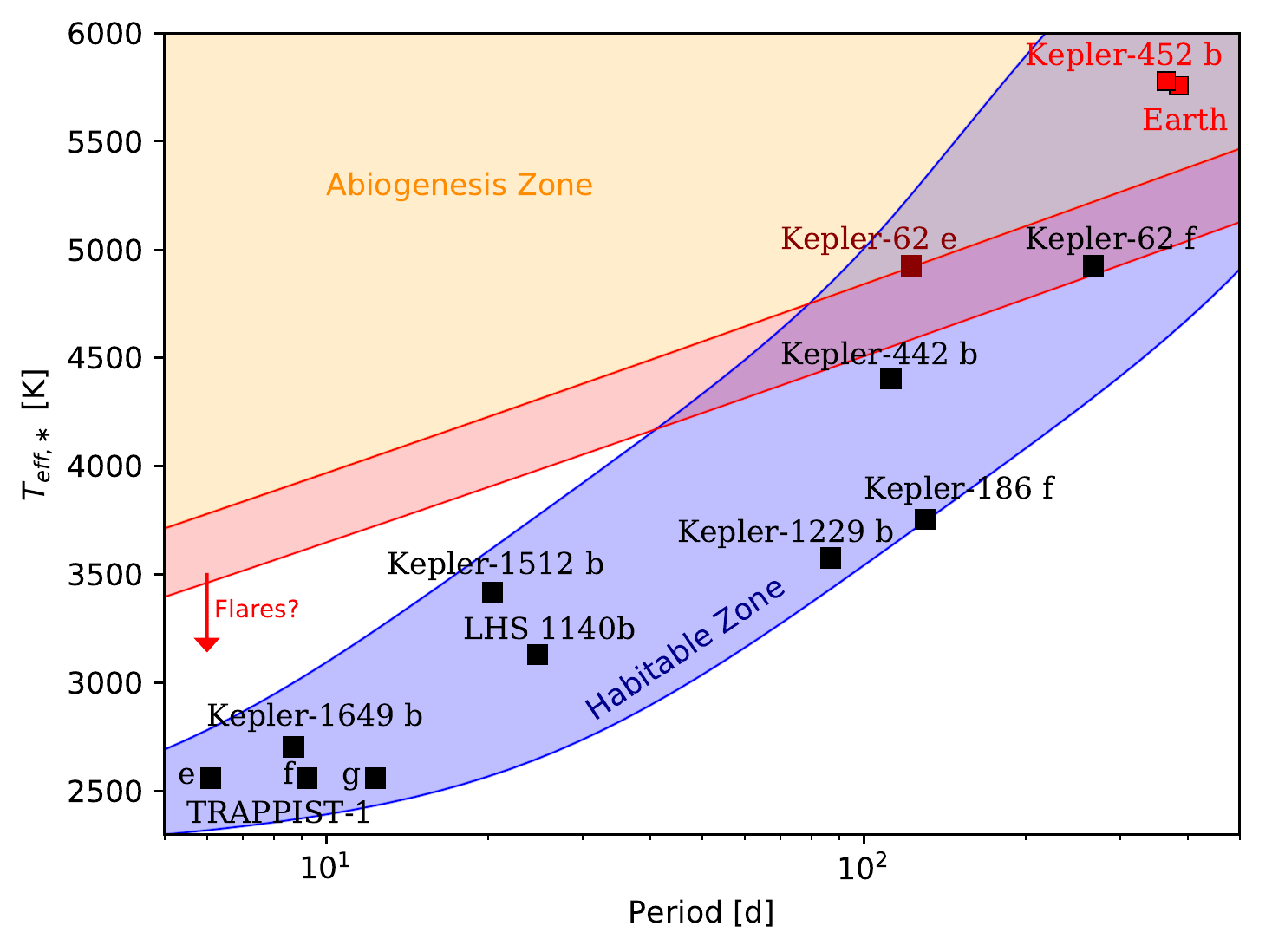}
\caption{{\bf Abiogenesis Zone.} A period-effective temperature diagram of confirmed exoplanets within the liquid water habitable zone (as well as Earth), taken from a catalog \cite{Kane2016,Morton2016,Angelo2017}, along with the TRAPPIST-1 planets \cite{Gillon2017}, and LHS 1140b \cite{Dittmann2017}, including only planets with $R_p \leq 1.4 R_{\oplus}$ (within error bars, and with Kepler 62e as an exception to highlight one rocky planet straddling the error bars of the abiogenesis zone). The `abiogenesis zone' indicates where the stellar UV flux is large enough to result in a 50\% yield of the photochemical product. The red region shows the propagated experimental error. The liquid water habitable zone (from \cite{Koppa2013},\cite{Koppa2014}) is also shown.}
\label{fig:habquiet}
\end{figure}

With a clear understanding of the temperature and UV flux where the SO$_3^{2-}$ chemistry can occur, we can now predict the planets on which this chemistry could occur. We take a catalog of all presently known potentially rocky exoplanets within the liquid water habitable zone of their host star. Using analog stellar spectra appropriate to the stellar types for the host stars in our catalog, we can show which exoplanets are primed for life and which cannot build up an appreciable prebiotic inventory. Fig. \ref{fig:habquiet} shows which of the liquid water habitable zone planets also lie within the abiogenesis zone. In order to select planets that are reasonably likely to be rocky, we impose a criterion on the planetary radius, $R_p \leq 1.4 R_{\oplus}$, where $R_{\oplus}$ is the radius of the Earth. The question of at what radius a planet is likely rocky is difficult to answer definitively. There have been many investigations into this question (e.g., \cite{Rogers2015,Wolfgang2015,Chen2017}).

Stars are not quiet, and several researchers have wondered whether the frequent flaring of young ultracool dwarves may be sufficient to initiate the photochemical production of the prebiotic inventory \cite{Rugheimer2015b,Ranjan2017c}. We explore this possibility. For our purposes, activity means variability in the wavelength range between 200 nm and 280 nm. This is primarily caused by flares and coronal mass ejections (CME's). Only a series of flares of sufficient energy over a period of time determined by the `dark chemistry' rates will result in a 50\% yield per step for the reactions along the way to the prebiotic inventory. 

To account for flaring, we consider a simple model. The dark chemistry is slowest at 0 \degree C, at which temperature high concentrations of SO$_3^{2-}$ and HCN are consumed at a rate determined by using our measured rate constants. If a sufficient number of flares occur such that a 50\% yield of the photochemical products is expected per step, then these planets meet the criterion for lying within their star's abiogenesis zone due to the stellar activity. We examine what fraction of stars cooler than K5 flare this frequently.

\begin{figure}[h!]
\includegraphics[width=\textwidth]{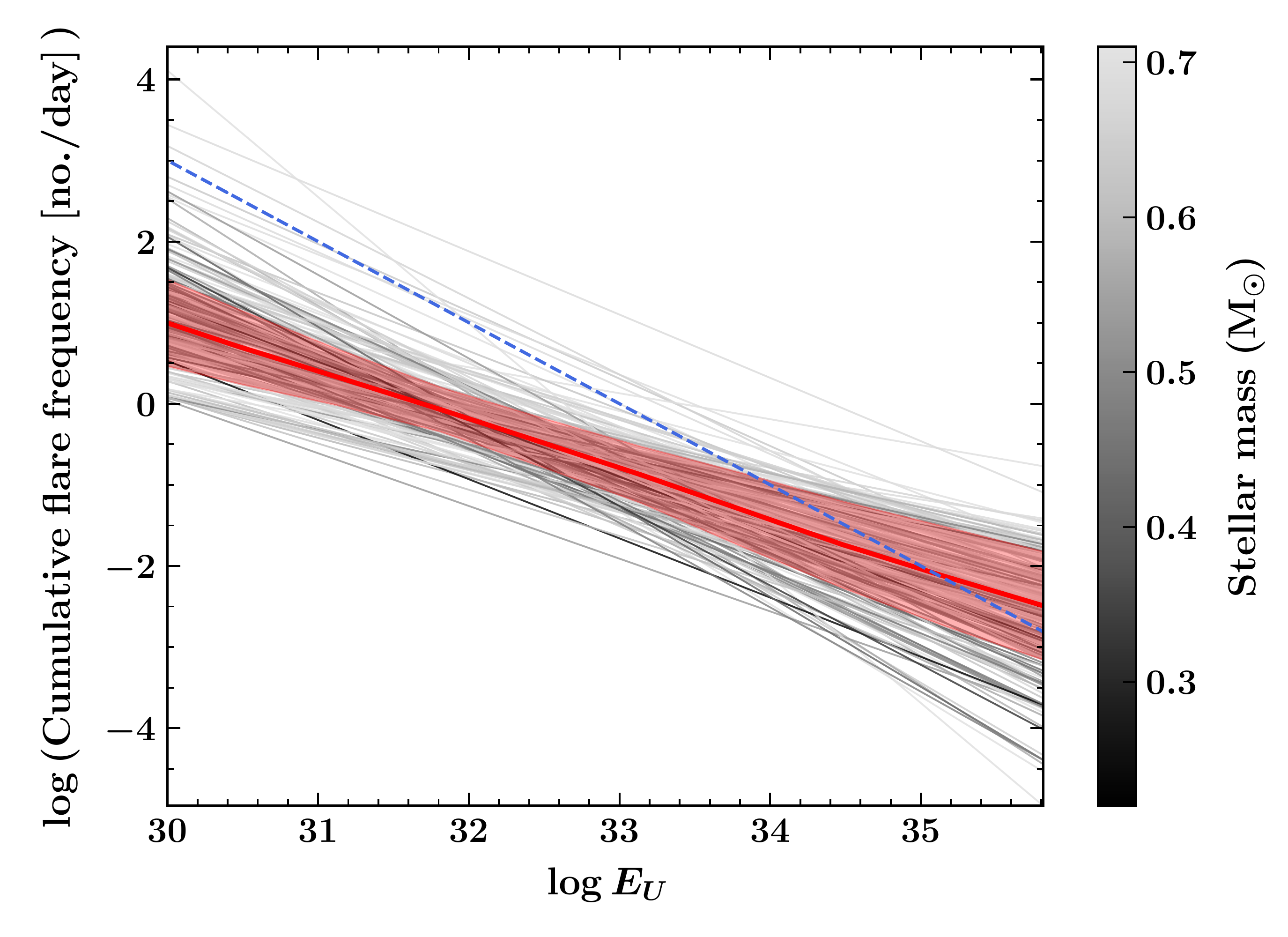}
\caption{{\bf Flare Frequencies.} The power-law fits for the cumulative flare frequencies above a particular U-band energy, $\log (E_U/{\rm 1 \, erg})$, given for a variety of Kepler stars from Davenport (2016,\cite{Davenport16}). The grayscale indicates stellar mass, the red line indicates the average stellar flare frequency for stars later than K5, and the blue dashed line indicates the lower limit of flaring needed: If the star's power-law flare frequency intersects the dotted blue line, the star is active enough for its planets to be considered within the abiogenesis zone. Flares with $E_U > 10^{35}$ are often extrapolations of flare frequencies beyond what has been observed, and flare frequencies may not follow the same power-law at such high energies \cite{Davenport16}.}
\label{fig:FFDs}
\end{figure}

Fig. \ref{fig:FFDs} shows our results using present flare statistics for cool stars \cite{Davenport16}. 
We find that $\sim$\,30\,$\%$ of stars cooler than K5, and $\sim$\,20\,$\%$ of the early M dwarfs, are active enough for the planets they host to be within their abiogenesis zones. The statistics given by Davenport (2016,\cite{Davenport16}) did not concentrate on high-energy flares, and is based on the Kepler sample, which does not include enough ultracool (cooler than M4) stars to provide very reliable flare statistics. Further analysis of the frequencies of energetic flares will be necessary for a more robust statistics about which exoplanets are primed for life due to the activity of their host stars. It would be especially relevant for flare rates to be determined for the stars included in Fig. \ref{fig:habquiet}.

Photons are also generated by the particles from the CME impinging on the upper atmosphere \cite{Air2016}, but even assuming that the total kinetic energy of every $> 1$ GeV proton from a large CME is converted into 200-280 nm photons, the flux would still be an order of magnitude too small to overcome the dark chemistry (Materials and Methods: Coronal Mass Ejections).

\section*{Discussion}
Using a known reliable pathway for photochemically building up the prebiotic inventory in large yields, we show that hotter stars serve as better engines for prebiotic chemistry. Investigating the race between light and dark bisulfite chemistry, we find, based on our requirement for $> 50\%$ yields, that even for the early Earth, the prebiotic inventory would need to be built up in places where the surface temperature is below $\sim 20$ \degree C. The HS$^-$ photodetachment is too slow to allow for the prebiotic inventory to build up on the Early Earth. As it turns out, however, HS$^-$ dark chemistry results in important amino acid precursors. Additionally, in the absence of UV light, HS$^-$ contributes toward a more efficient pathway to form pyrimidine nucleosides \cite{Xu2016}. Since UV flux of the young sun would not have been sufficient to drive the HS$^-$ photochemistry, the HS$^-$ chemistry behaves as though it is taking place in the absence of UV light. With both HS$^-$ and bisulfite present, this provides both varieties of chemistry (bisulfite light chemistry and HS$^-$ dark chemistry) simultaneously: the best of both worlds.

Because of the efficiency of the bisulfite photochemistry, rocky planets within the liquid water habitable zones of K dwarfs can also lie within the abiogenesis zone, so long as the temperature is very close to 0 \degree C. We applied our results to a catalog of potentially rocky exoplanets within the liquid water habitable zones of their host stars. For gas giants within the liquid water habitable zone, there is a tantalizing possibility that some of their larger moons may be primed for life \cite{Forgan2013}.

The abiogenesis zone we define need not overlap the liquid water habitable zone. The liquid water habitable zone identifies those planets that are a sufficient distance from their host star for liquid water to exist stably over a large fraction of their surfaces. In the scenario we consider, the building blocks of life could have been accumulated very rapidly compared to geological timescales, in a local transient environment, for which liquid water could be present outside the liquid water habitable zone. Such local and transient occurrences of these building blocks would almost certainly be undetectable. The liquid water habitable zone helpfully identifies where life could be sufficiently abundant to be detectable.

For main sequence stars cooler than K5 dwarfs, the quiescent stellar flux is too low for the planets within their habitable zones to also lie within their abiogenesis zones. Planets within the habitable zones of {\bf quiet} ultracool dwarfs may be able to house life, but life could not presently originate as a result of photochemistry on these worlds, although it possibly could have done in the past, if these stars emitted much more strongly in the UV before they entered into the main sequence, or if they had been much more active in the past. Our results are only valid for the stars as they are now. There is a decent chance that, for the most active M dwarfs, flares could be sequentially timed during intermediate reactions along the chemical pathway to build up the prebiotic inventory.

It turns out that stellar activity is not always bad for life, but may in fact be the only pathway to starting life on planets around ultracool stars. If the activity of an ultracool dwarf decreases as it ages, this might allow there to be sufficient 200-280 nm light from flares to initiate life soon after the star's formation, but as the star gets older, the 200-280 nm light could decrease to a low enough level so that resulting RNA will not be damaged as frequently, although flares may still pose a problem for the stability of RNA strands on these planets.

There are other theories about how the prebiotic inventory could be generated, either on the surface of the planet \cite{Cleaves2008,Ruiz2014} or on interstellar ices \cite{Bernstein2002}. These other theories may provide plausible pathways to form a variety of amino acids, or some of the nucleobases, and so may help provide some of the building blocks of life, if somehow the relevant species can be selectively concentrated. But none of these theories has offered prebiotically plausible chemical pathways to high yields of a range of nucleotides, amino acids and lipid precursors from prebiotically plausible starting conditions, which remains a necessary first step for the origin of life. For hypothetical alternative theories, if there is a clear link between their chemical pathways and observable physical conditions of the planet or system, a similar analysis can be carried out and abiogenesis zones can be identified for them.

What do our results say about looking for biosignatures on rocky planets around ultracool dwarfs? This is an important consideration, given that ultracool dwarfs are likely the only possible candidates for this search within the next decade or so. Present flare statistics for cool stars suggests that as many as 20\% of M dwarfs may flare often enough to drive the prebiotic photochemistry on rocky planets within their habitable zones, assuming that the atmospheres of these planets can survive such extreme activity. What is most relevant for biosignature detection would be the flare rates for M dwarfs early in their evolution, since we will want to search for biosignatures on planets where life has existed long enough to change the planet's surface and atmosphere in detectable ways. It is important to note that the liquid water habitable zones of M-dwarfs evolve inward as they age \cite{Rushby2013}. If atmospheres are observed for a large fraction of rocky planets around active ultracool stars, then it may well be the case that life is more likely to have originated in systems with the most active stars, all else being equal.

The search for biosignatures on planets around quiet M-dwarfs remains worthwhile, in some senses even more now than before our investigation. If definitive biosignatures were discovered within the atmospheres of multiple rocky planets around quiet ultracool dwarfs, at the very least, it would suggest that the mechanism by which Earth-like life could originate is not universal.

\section*{Materials and Methods}

\subsection*{Experimental Design}
\label{sup:lamp-flux}

To determine the spectral irradiance of the lamps used in the photochemistry experiments presented here we used a calibrated OceanOptics FLAME-S UV-Vis spectrometer with a UV cosine-corrected irradiance probe (connected with a 2 m length of 600 $\mu$m UV fiber).  The probe collects radiant light with a Lambertian $180^{\circ}$ field of view. The spectrometer is calibrated using the OceanOptics DH-3-CAL, DH-3-CAL-EXT Calibrated Deuterium and Halogen light source, valid from 200 nm to 700 nm.  The measurements were conducted within a Rayonet UV reactor which is set-up with 16 Hg lamps.  A diagram illustrating the general experimental set-up is shown in Fig. S1.

For this study we only consider the flux in the emission line at 254 nm which contributes to approximately 90\% of the total spectral irradiance over the 200-400 nm wavelength region.

The total UV flux received by the sample is calculated by first taking a measurement with the spectrometer at the approximate centre of the quartz cuvette, 1cm from the nearest lamp.  Every lamp/cuvette position was measured in the same way to verify that the irradiance from each lamp is approximately equal.  Given that the spectrometer probe only has a 180 degree field of view and that the cuvette effectively sees all the lamps, the irradiance received by one cuvette is then calculated by measuring the flux received by the cuvette by the nearest lamp only (with the other lamps switched off) and then using this measurement to calculate the contribution of all the other lamps based on their distance away from the centre of the cuvette.

The measured spectrum over the wavelength range 200-400nm, with the sample flux calculated as previously described, is shown in Fig. S2.  The best-fit Gaussian profile for the emission line at 254 nm is shown by the dotted line on Fig. S2 and is described by Eq. \ref{eq:lamp_fit}.  The total UV flux received by a sample is then the area under this curve multiplied by the total surface area of the cuvette and the duration time the sample is illuminated for.

\begin{equation}
F_{\lambda,{\rm lamp}} = \dfrac{F_0 \Delta \lambda}{\sigma \, \sqrt{2\pi}}e^{-(\lambda-\lambda_0)^2/(2\sigma^2)}
\label{eq:lamp_fit}
\end{equation}
where $F_0 = 2.54 \times 10^{15}$ cm$^{-2}$ s$^{-1}$ \AA$^{-1}$, $\lambda_0$ = 254 nm, $\sigma$ = 0.6 nm, $\Delta \lambda = 1$ nm.

\subsection*{Model}
\label{sup:star-flux}

To determine the UV spectral irradiance on the surface of an exoplanet, we have to start with the irradiance of the star from space. In order to do so, we take the MUSCLES version 2.1 spectra of Proxima Cen, GJ832, and $\epsilon$ Eri \cite{France2016,Youngblood2016,Loyd2016}. For the Early Earth, we take a solar analogue spectrum of $\kappa 1$ Ceti \cite{Ribas2010}. For AD Leo, we use the spectrum from Segura et al. (2005, \cite{Segura2005}, incomplete for the FUV). In all cases, these spectra are taken from Earth and are in units of erg cm$^{-2}$ s$^{-1}$ \AA$^{-1}$, which flux we will call $S_{\lambda,\oplus}$. From here, we need to get the flux at the very top of our exoplanet's atmosphere $S_{\lambda,p}$, in units of photons cm$^{-2}$ s$^{-1}$ \AA$^{-1}$, and to do so, we take:
\begin{equation}
S_{\lambda,p} = \dfrac{hc}{\lambda}\dfrac{a^2}{d^2}S_{\lambda,\oplus},
\end{equation}
where $d$ is the distance from the star to the observer, and $a$ is the semimajor axis of the hypothetical planet, always chosen so that the bolometric flux from the star is identical to the bolometric flux from the sun, following earlier work \cite{Rugheimer2013}. Here $h$ is Planck's constant, $c$ is the speed of light and $\lambda$ [nm] is the wavelength of the light. $S_{\lambda,p}$ is shown in Fig. S3. 

We take these spectra, and apply them as the top boundary condition to the ARGO photochemistry/diffusion code \cite{Rimmer2016}. This code solves the set of photochemical kinetics equations for a moving parcel of atmospheric gas:
\begin{equation}
\dfrac{\partial n_i}{\partial t} = P_i - L_i - \dfrac{\partial \Phi_i}{\partial z},
\end{equation}
for an atmosphere of bulk composition 80\% $N_2$, 20\% $CO_2$, and other species  from the bottom of the atmosphere in trace amounts. This is similar to the presumed atmosphere of the Early Earth \cite{Kasting1993}. Given that volatile acquisition is unsolved for our own solar system, and that the secondary atmosphere is determined primarily by the outgassing pressure and C/O ratio, CO$_2$ is expected to be a dominant component of the atmosphere of a rocky planet of approximately Earth mass and with a crustal and upper mantle C/O ratio of $\lesssim 0.8$ \cite{Gaillard2014}\cite{Hu2014}. Why the Earth has its present quantity of atmospheric nitrogen is a mystery, and at present we have no alternative recourse but to assume similar amounts of N$_2$ in the atmospheres of rocky exoplanets. Since N$_2$ does not absorb   in UV between 200-280 nm \cite{Ranjan2016}, its presence or absence by itself should not significantly affect these results.  Here, $n_i$ [cm$^{-3}$] is the number density of species $i$, where $i = 1,...,I_s$, $I_s$ being the total number of species. $P_i$ [cm$^{-3}$ s$^{-1}$] and $L_i$ [cm$^{-3}$ s$^{-1}$] represent the production and loss rates, which are determined by using the STAND chemical network \cite{Rimmer2016}. The quantity $\Phi_i$ is the flux of the species $i$ into and out of the parcel when it is at height $z$. As described in \cite{Rimmer2016}, the parcel is used to give a chemical profile of the atmosphere, and UV photo-transport is calculated in order to determine the efficiency of the photochemistry at each parcel height. At the very end, the surface actinic flux is determined by taking $S_{\lambda,p} = F_{\lambda}(z_{\rm top})$, where $z_{\rm top}$ [cm] is the height of the top of the atmosphere, above which the UV light in consideration is virtually unattenuated. The surface atmospheric flux is then:
\begin{equation}
F _{\lambda}(z=0) = S_{\lambda,p}e^{-\tau(\lambda,z=0)/\mu_0} + F_d,
\end{equation}
where $\tau$ is the optical depth determined using the chemical profile, and is unitless, $F_d$ is the diffusive actinic flux determined using the $\delta$-Eddington 2-stream method \cite{Toon1989},and $\mu_0$ is the cosine of the stellar zenith angle, which we set $=1$ to achieve the maximum surface flux, the best case scenario for our light chemistry. The surface UV actinic fluxes for the hypothetical or real habitable zone planets around the young Sun, Proxima Cen, GJ832 and $\epsilon$ Eri are shown in the main paper.

In order to compare our model to data, we used the OceanOptics spectrometer and measured the sun outside the Laboratory of Molecular Biology in Cambridge on 24 February 2017 at 15:00. We used the NOAA Solar Calculator (\url{https://www.esrl.noaa.gov/gmd/grad/solcalc/azel.html}) in order to determine the cosine of the zenith angle, and applied our model with an 80\% N$_2$, 20\% O$_2$ atmosphere, with some trace constituents. The comparison between our model and the data is given in Fig. S4.

\subsection*{Rate Constants}
\label{sup:kinetics}
In this section, we consider the kinetics for the reactions discussed in the Supplementary Material. The kinetics two 'dark' reactions we consider (bimolecular reactions in the absence of 254 nm light):
\begin{equation}
\schemestart
\chemfig{KCN} + \chemfig{Na_2SO_3} \arrow \chemfig{HSO_3-[:30](-[2,0.65]NH_2)-[:-30]\chemabove{SO_3}{\scriptstyle\ominus}}
\schemestop
\end{equation}
\vspace{5mm}
\begin{equation}
\schemestart
\chemfig{NaHS}$\cdot$\chemfig{XH_2O} + \chemfig{OH-[:30,0.65]-[:-30,0.65]CN} \arrow \chemfig{HO-[:30]-[:-30](-[:30]NH_2)=[:-90,0.65]S}
\schemestop
\end{equation}
follow the rate equation mediated by the rate constant $k_r$ [L mol$^-1$ s$^-1$]:
\begin{equation}
\dfrac{dx}{dt} = k(c_0 - x)^2
\end{equation}
for stoicheometrically balanced initial concentrations of reactants ($c_0$ [mol/L]), where $x$ is the concentration of the adduct. The solution to this equation, considering the boundary condition where $x(0) = 0$, is:
\begin{equation}
x(t) = c_0 - \dfrac{1}{\frac{1}{c_0} + k_rt}.
\end{equation}
We use this solution, along with the NMR measurements of the products discussed in Section 2.3, in order to solve for $k$. The NMR measurements were taken for mixtures of KCN $+$ Na$_2$SO$_3$, as well as NaHS$\cdot x$H$_2$O $+$ C$_2$H$_3$NO, at different temperatures, but with the same initial $c_0$. We refer to the KCN $+$ Na$_2$SO$_3$ reaction by the reaction that takes place from this starting material in water: SO$_3^{2-}$ + HCN, and NaHS$\cdot x$H$_2$O $+$ C$_2$H$_3$NO by HS$^-$ + C$_2$H$_3$NO. Either by measuring the product or, where possible, both product and reactant, we determine the concentration of the adduct, $x$, and use this to solve for $kt$. For some temperatures, we make multiple measurements of $k$, to estimate the variance. The errors in these measurements are dominated by errors in weighing the reactants and, in the case of HS$^{-}$, the loss of the bound H$_2$O and the escape of the reactant in the form of H$_2$S.

The resulting concentrations, times and rate coefficients are shown for each measurement in Tables S1 and S2. Arrhenius plots for both bimolecular reactions can be produced (Fig. S5). These are fit using the SciPy routine, assuming the standard Arrhenius form for the rate coefficient, in this case:
\begin{equation}
\log k_r = \log \Bigg(\dfrac{\alpha}{1 \, \text{mol$^{-1}$ L s$^{-1}$}}\Bigg) - \dfrac{E_a}{T},
\end{equation}
where $\alpha$ is the prefactor, the rate coefficient for the reaction at a hypothetical infinite temperature, $E_a$ [K] is the activation energy, and $T$ [K] is the temperature of the solution. For the two reactions, we achieve the results:
\begin{align}
\log \alpha({\rm SO_3^{2-} \, + \, HCN}) &= 18.7\text{$\pm$ 0.32} \; {\rm mol^{-1} \, L \, s^{-1}}, \label{eqn:R1}\\
E_a({\rm SO_3^{2-} \, + \, HCN}) &= \text{9490 $\pm$ 106} \; {\rm K}, \label{eqn:R2}\\
\log \alpha({\rm HS^- \, + \, C_2H_3NO}) &= 12.2 \pm 0.3 \; {\rm mol^{-1} \, L \, s^{-1}}, \label{eqn:R3}\\
E_a({\rm HS^- \, + \, C_2H_3NO}) &= \text{7480 $\pm$ 130} \; {\rm K}. \label{eqn:R4}
\end{align}
Which are placed in more natural units within the main text.

For the light chemistry (photodetachment reactions under 254 nm light, as well as the bimolecular reactions), we consider the reactions:
\begin{equation}
\schemestart
\chemfig{KCN} + \chemfig{Na_2SO_3} \arrow{->[$h\nu$]} \chemfig{OH-[:30,0.65]-[:-30,0.65]CN}
\schemestop
\end{equation}
\vspace{5mm}
\begin{equation}
\schemestart
\chemfig{NaHS}$\cdot$\chemfig{XH_2O} + \chemfig{HO-[:30,0.65]-[:-30,0.65]CN} \arrow{->[$h\nu$]} \chemfig{[:-90,0.5](-[:30,0.5]OH)-=[:-30,0.5]O}
\schemestop
\end{equation}
It should be noted that, in the case of KCN + Na$_2$SO$_3$, many of the photochemical products are intermediates toward glycolonitrile, such as aminomethanesulfonate and azanediyldimethanesulfonate, which will proceed under the conditions we consider (pH 7) almost completely to glycolonitrile via reactions that proceed without the UV light \cite{Xu2018}. As explained above, the abundance of these intermediates is included and treated here as though the intermediates were glycolonitrile. 

The rate coefficients for these reactions are determined in terms of the consumption of the reactant, $y$ [mol/L], mediated by both the photodetachment rate $k_{\lambda}$ [s$^{-1}$] and the bimolecular reaction $k_r$ [L mol$^{-1}$ s$^{-1}$], same as above:
\begin{equation}
\dfrac{dy}{dt} = -k_r y^2 - k_{\lambda} y,
\end{equation}
which can be integrated directly to achieve:
\begin{equation}
-k_{\lambda}t = \log \Bigg(\dfrac{y}{y + \frac{k_{\lambda}}{k_r}}\Bigg) + C,
\end{equation}
where $C$ is the constant of integration. This is left as the following equation, with $C$ solved for the boundary condition $y(0) = c_0$:
\begin{equation}
\dfrac{y}{y + \frac{k_{\lambda}}{k_r}} = \Bigg(\dfrac{c_0}{c_0 + \frac{k_{\lambda}}{k_r}}\Bigg)e^{-k_{\lambda}t}.
\label{eqn:light-solution}
\end{equation}
This solution is used to numerically solve for $k_{\lambda}$ under different fluxes for the SO$_3^{2-}$ chemistry, by placing the mixture within the reactor when all the lamps are on, by having half the lights on and placing the mixture near one of the lights that is on, and by having half the lights on and placing the mixture near one of the lights that is off. The actinic flux for these three cases was calculated in the Experimental Design. When $k_{\lambda}$ is solved, we make the assumption, which has weak empirical support \cite{Sauer2004,Todd2018}, that the reaction cross-section for photodetachment, $\sigma_{\lambda}$ [cm$^2$], is constant over the range 200 nm - 280 nm. Given the actinic flux, $F_{\lambda}$ [photons cm$^{-2}$ s$^{-1}$ \AA$^{-1}$] and the equation for the photochemical rate constant:
\begin{equation}
k_{\lambda} = \int_{\rm 200 \, nm}^{\rm 280 \, nm} F_{\lambda} \sigma_{\lambda} \, d\lambda,
\label{eq:photo-sigma}
\end{equation}
and, since we are assuming $\sigma_{\lambda}$ is constant over the range of $\lambda$, so it comes out in front of the integral, and the integral of $F_{\lambda}$ is simply the ERF prefactor for the Gaussian fit to the peak 254 nm lamp flux (Experimental Design). In this way, we take the product of the irradiation, $z$ [mol/L] and the product of the dark chemistry occurring alongside the light chemistry, $x$ [mol/L], and set $y = c_0 - x - z$. We then numerically solve for $k_{\nu}t$ from the above solution, divide by the time under which the mixture has been irradiated, and divide again by $\int F_{\lambda} \, d\lambda$ to determine that:
\begin{equation}
\sigma_{\lambda}({\rm SO_3^{2-} \, + \, HCN}) = (1.5 \pm 0.3) \times 10^{-20} \; {\rm cm^2}.
\label{eqn:photo-so32-}
\end{equation}
We were unable to perform the same analysis for HS$^-$, becuase HS$^-$ is not as soluable, and escapes the mixture and oxidizes, which would require us to incorporate other loss terms in our kinetics equation. Instead, we have taken a different approach, as discussed above, and saturated our mixture with HS$^-$ compared to the glycolonitrile. This modifies the rate equation to:
\begin{equation}
\dfrac{dy}{dt} = -k_r y_0y - k_{\lambda} y,
\end{equation}
where $y$ in this case is the concentration of the glycolonitrile, and $y_0$ is the HS$^-$, which is effectively constant compared to the glycolonitrile. This is a linear differential equation with the solution:
\begin{equation}
y(t) = c_0 e^{-(k_ry_0 + k_{\lambda})t}.
\end{equation}
We use the `dark chemistry' values for $k_r$, use our initial concentration of HS$^-$ for $y_0$, and the initial concentration of glycolontirile for $c_0$, and solve for $k_{\lambda}$. Then, with the same assumption of the constant photochemical cross-section, we apply $k_{\lambda}$ to Eq. (\ref{eq:photo-sigma}) and find:
\begin{equation}
\sigma_{\lambda}({\rm HS^- \, + \, C_2H_3NO}) = (2.0 \pm 0.4) \times 10^{-23} \; {\rm cm^2}.
\label{eqn:photo-hs-}
\end{equation}
The values of $\sigma_{\lambda}$ from Eq's (\ref{eqn:photo-so32-}),(\ref{eqn:photo-hs-}) are the same value to within the error bars.

To quantify the competition between ``light'' and ``dark'' chemistry, we compare the half-lives of the two reactions. In the light, this is simply:
\begin{align}
t_{\rm light} &= \dfrac{\log 2}{\int_{\rm 200 \, nm}^{\rm 280 \, nm} F_{\lambda} \sigma_{\lambda} \, d\lambda}, \\
&= \dfrac{\log 2}{\sigma_{\lambda}  \overline{F_{\lambda}} \Delta \lambda},
\end{align}
where $\overline{F_{\lambda}}$ [cm$^{-2}$ s$^{-1}$ \AA$^{-1}$] is the average actinic flux at the surface between 200 nm and 280 nm, and $\Delta \lambda = 80$ nm is the relevant wavelength range.

The halflife for the dark reaction is:
\begin{equation}
t_{\rm dark} = \dfrac{1}{c_0k_r}.
\end{equation}
We take the most favorable temperature for the reactions, that the surface temperature is 0 $^{\circ}$C. We assume a high initial concentration of reactants, $c_0 = 1$ M. Although this makes the dark reaction fast, decreasing this value significantly will frustrate subsequent bimolecular reactions needed for the pyrimidine synthesis. The effect of decreasing the concentration has a linear effect on the rate for the reaction in the dark, but a quadratic effect on the rates of these subsequent bimolecular reactions.

We set the $t_{\rm light} = t_{\rm dark}$, and solve for $\overline{F_\lambda}$:
\begin{equation}
\overline{F_\lambda} = \dfrac{c_0 k_r \log 2}{\sigma_\lambda \Delta \lambda}.
\end{equation}
Propagating the errors through this equation gives us:
\begin{equation}
\Delta\overline{F_\lambda} = \dfrac{c_0 \log 2}{\Delta \lambda \sigma_\lambda}\sqrt{\big(\Delta k_r\big)^2 + \dfrac{k_r^2}{\sigma_\lambda^2}\big(\Delta \sigma_{\lambda}\big)^2},
\end{equation}
where $\Delta k_r$ is the error in the bimolecular rate constant, from Eq's. (\ref{eqn:R1})-(\ref{eqn:R4}), and $\Delta \sigma_\lambda$ is the error in the photochemical cross-section, from Eq's (\ref{eqn:photo-so32-}),(\ref{eqn:photo-hs-}). In order for the light chemistry to compete with the dark chemistry at 0 \degree C, we find that for SO$_3^{2-}$: 
\begin{equation}
F_{\lambda} = (6.8 \pm 3.6) \times 10^9 \; {\rm cm^{-2} \, s^{-1} \, \AA^{-1}},
\end{equation}
and for HS$^-$:
\begin{equation}
F_{\lambda} = (1.6 \pm 0.4) \times 10^{12} \; {\rm cm^{-2} \, s^{-1} \, \AA^{-1}},
\end{equation}
integrated over 200 - 280 nm. For SO$_3^{2-}$, this is the rough equivalent of the integrated quiescent flux of a K5 dwarf ($T_{\rm eff} \approx 4400$ K).

\subsection*{Flare Rates}
\label{sup:flares}

When we model the flares, we consider as our sole template the spectrum of the 1985 April 12 AD Leo flare \cite{Hawley1991}, which had an energy in the U-band of $E_U = 10^{34}$ erg. We then change the spectrum as a function of time following the spectral evolution given by \cite{Segura2010}. This suffices to describe the time-evolution of the UV flux for a $10^{34}$ erg flare, which we assume here can be applied in the same way to all stars. We scale the peak flux and duration using flare statistics from Davenport (\cite{Hawley14}, their Fig. 10). We then pass it through the atmosphere as describe (Experimental Design). This is how we produce Fig. S6. Our treatment of the flares, however, can rest more generally on the flare statistics, without worrying as much about the specifics of the AD Leo flare.

The total number of photons necessary within the lifetime of the reactant for a 50 \% yield of the photochemical product is:
\begin{align}
n_{\gamma,r} &= \dfrac{1}{\sigma_{\lambda}\Delta \lambda}, \notag\\
&= 8.3 \times 10^{16} \; \text{cm$^{-2}$ \AA$^{-1}$}, \label{eqn:photon-rate-flare}
\end{align}
using the value of $\sigma_{\lambda}$ from Eq. (\ref{eqn:photo-so32-}).

We note that the spectrum of the AD Leo flare is relatively flat at wavelengths shorter than the U band, and so the total number of photons cm$^{-2}$ \AA$^{-1}$ will be the same between 200 nm and 280 nm as it is within the U band. As such, we can estimate the total fluence (the total number of photons deposited during the length of the flare) at the planet's surface for photons between 200 nm and 280 nm to be linearly proportional to the flare energy in the U band, or:
\begin{equation}
n_{\gamma,f} = 2.4 \times 10^{16} \, \text{cm$^{-2}$ \AA$^{-1}$} \; \Bigg(\dfrac{E_U}{\text{$10^{34}$ erg}}\Bigg).
\label{eqn:energy-rate-flare}
\end{equation}
This equation relies on correlations between the fluence of the flare and its energy in the U-band. This can vary considerably, leading to large errors.
For the reaction to exceed 50\% conversion, $n_{\gamma,f} \geq n_{\gamma,r}$.

Flare frequencies follow a power-law distribution \cite{Hilton2011,Hawley14}, such that:
\begin{equation}
\log \nu = \log \alpha + \beta \log E_U,
\end{equation}
where $\nu$ [day$^{-1}$] is the cumulative flare frequency for all flares of U-band energy greater than $E_U$, and $\alpha$ and $\beta$ are coefficients for the power-law fit. This equation gives us our constraint on which power-laws are necessary for building up the prebiotic inventory, solving for $\nu$ after applying Eq's (\ref{eqn:photon-rate-flare}) and (\ref{eqn:energy-rate-flare}), we find:
\begin{equation}
\nu = \dfrac{8 \times 10^{27} \; {\rm erg/s}}{E_U}.
\label{eqn:flare-frequency}
\end{equation}
For the power laws published by Eric Hilton \cite{Hilton2011}, which come from targeted ground-based observations of $\sim$25 stars, only those for the very Active M3-M5 dwarfs qualify.

The \kepler\ mission, while not specifically targeting M-dwarfs, still observed many of these low-mass stars that fell within its field of view. Multiple studies have analyzed flares in \kepler\ data 
\cite{Hawley14,Davenport14,Davenport16}. A homogeneous search for stellar flares was performed by Davenport (2016; \cite{Davenport16}) using all available \kepler\ light curves. \kepler\ data spans 3.5 years, which provides a suitable baseline from which to estimate flare rates. Flares were detected on $\sim$2\% of all \kepler\ stars ($\sim$4000 in total) and flare frequency distributions (FFDs) computed. 

We take the Davenport FFDs and calculate the percentage whose FFD implies they produce enough flares to satisfy Eq. (\ref{eqn:flare-frequency}). To do this we: 1. convert energy in the \kepler\ band to the $U$ band following the relation presented in \cite{Hawley14}, i.e. $E_{U} = 0.65 E_{K_{p}}$; 2. remove eclipsing binaries through cross-matching with the Kepler Eclipsing Binary Catalogue V3 \cite{Kirk16}; and 3. require a system to show a cumulative flare frequency of at least 1 day$^{-1}$ at $E_{U} \geqslant 1\times 10^{30}$ erg, which removes non-active stars and hence lowers the chances of extrapolating the FFDs.

Fig. \ref{fig:FFDs} shows the FFDs for all stars passing our selection criteria. We evaluate the stars whose power-law flare rates cross the power-law of Eq. (\ref{eqn:flare-frequency}) as potentially hosting planets within their abiogenesis zones. We split the flare stars into different mass bins and report the percentages of sufficiently active stars in Table S3. We find that the highest flare rates are for stars between $M=0.6-0.7$\,$M_{\odot}$.

Although we made our best efforts to remove eclipsing binaries, for which it is more difficult to identify flares, for which flaring rates might be very different, and for which the question of habitability is more controversial, we have not been able to remove all binaries. Namely, we have not been able to remove the non-eclipsing binaries from our sample. Furthermore, the statistics from Davenport's sample often require extrapolations, sometimes extreme, from the energies at which the flares were observed to the higher-energy flares. Extrapolating this much off of a power law is always fraught with risk, especially when Davenport has pointed out that it is common to find breaks in the power laws for flare rates (\cite{Davenport16}, their Fig. 6). Extrapolating to low energies is also unreliable, because Davenport's methods for detecting flares may not be sensitive to flares below a certain energy.

However, since in every example given, the power law breaks steeper, our statistics will provide at the very least a reliable upper limit for the number of stars earlier than M4, for which flares are both energetic enough and frequent enough to produce the prebiotic inventory (see Table S3). We note that previous studies found that flares are more frequent on mid-M dwarfs (M3-M5) than on earlier-M dwarfs (M0-M2), but typically have lower energies.
Due to all these uncertainties, and contradictory findings, it will be important to reanalyze flare rates from data and possibly new observations, concentrating on very energetic flares, with energies above $10^{34}$ erg. It will be especially useful to perform these statistics for those stars from the Kane catalog \cite{Kane2016} and host stars for rocky planets potentially within their habitable zones.

\subsection*{Coronal Mass Ejections}
\label{sup:CME}

In order to estimate the effect of energetic particles from a Coronal Mass Ejection (CME) on the UV surface flux, we consider only those particles that impinge on the upper atmosphere effectively, without being deflected by the planetary magnetic field, and which will produce particle showers upon impact with the atmosphere. This restricts us mostly to CME protons of energy greater than 1 GeV. We use for our analysis the same energy distribution of CME protons as \cite{Air2016}, or:
\begin{equation}
j(E) = 1.26 \times 10^6 \, \text{protons cm$^{-2}$ GeV$^{-1}$} \; \Bigg(\dfrac{E}{\rm 1 \; GeV}\Bigg)^{\!\!-2.15},
\end{equation}
where $E$ [GeV] is the proton energy. We consider that a CME of this strength impinges on the atmosphere every 5.8 minutes ($\tau_{\rm CME} = 350$ s), based on the CME event rates estimated by \cite{Air2016}. We make the assumption that the energy of every CME proton is converted entirely into photons of wavelength $\lambda_0 = 280$ nm. This assumption is unphysical, but it provides a hard upper limit to the impact of the CME particles on the UV photochemistry. In this case, the number of photons produced by a cosmic ray of energy $E$ is:
\begin{equation}
N_{\gamma} = \dfrac{E \lambda_0}{hc},
\end{equation}
where $h$ is Planck's constant and $c$ is the speed of light. The photochemical rate constant in this case will be the total number of 280 nm photons times the cross-section at 280 nm or, for SO$_3^{2-}$, where $\sigma_{\lambda} = 1.5 \times 10^{-21}$ cm$^2$:
\begin{equation}
k_{\lambda,{\rm CME}} = \dfrac{\lambda_0 \sigma_{\lambda}}{hc\tau_{\rm CME}}\int_{\rm 1 \, GeV}^{\infty} E j(E) \, dE.
\end{equation}
The solution is that $k_{\lambda,{\rm CME}} = 8.1 \times 10^{-9}$ s, giving a half-life for the light chemistry due to CME's of about 2.7 years. This is too slow to produce the prebiotic inventory.

\section*{Supplementary Materials}
Fig. S1. The UV Reactor\\
Fig. S2. Reactor Emission\\
Fig. S3. Stellar Spectra\\
Fig. S4. Surface Spectral Irradiance at Cambridge\\
Fig. S5. Ahrennius Plots\\
Fig. S6. Flare Spectra
Table S1. Values from HCN + SO$_3^{2-}$ Experiments\\
Table S2. Values from HCN + HS$^-$ Experiments\\
Table S3. Percentages of stars in dierent spectral type ranges that meet the criterion for activity given by Eq. (38) (defined here as "Active" stars).

\bibliographystyle{ScienceAdvances}

\begin{thebibliography}{10}
\bibitem{Kane2016}
S.~R. {Kane}, M.~L. {Hill}, J.~F. {Kasting}, R.~K. {Kopparapu}, E.~V.
  {Quintana}, T.~{Barclay}, N.~M. {Batalha}, W.~J. {Borucki}, D.~R. {Ciardi},
  N.~{Haghighipour}, N.~R. {Hinkel}, L.~{Kaltenegger}, F.~{Selsis},
  G.~{Torres}, {A Catalog of Kepler Habitable Zone Exoplanet Candidates}.
\newblock {\it ApJ\/} {\bf 830}, 1 (2016).

\bibitem{Anglada2016}
G.~{Anglada-Escud{\'e}}, P.~J. {Amado}, J.~{Barnes}, Z.~M. {Berdi{\~n}as},
  R.~P. {Butler}, G.~A.~L. {Coleman}, I.~{de La Cueva}, S.~{Dreizler},
  M.~{Endl}, B.~{Giesers}, S.~V. {Jeffers}, J.~S. {Jenkins}, H.~R.~A. {Jones},
  M.~{Kiraga}, M.~{K{\"u}rster}, M.~J. {L{\'o}pez-Gonz{\'a}lez}, C.~J.
  {Marvin}, N.~{Morales}, J.~{Morin}, R.~P. {Nelson}, J.~L. {Ortiz}, A.~{Ofir},
  S.-J. {Paardekooper}, A.~{Reiners}, E.~{Rodr{\'{\i}}guez},
  C.~{Rodr{\'{\i}}guez-L{\'o}pez}, L.~F. {Sarmiento}, J.~P. {Strachan},
  Y.~{Tsapras}, M.~{Tuomi}, M.~{Zechmeister}, {A terrestrial planet candidate
  in a temperate orbit around Proxima Centauri}.
\newblock {\it Nature\/} {\bf 536}, 437-440 (2016).

\bibitem{Gillon2017}
M.~{Gillon}, A.~H.~M.~J. {Triaud}, B.-O. {Demory}, E.~{Jehin}, E.~{Agol}, K.~M.
  {Deck}, S.~M. {Lederer}, J.~{de Wit}, A.~{Burdanov}, J.~G. {Ingalls},
  E.~{Bolmont}, J.~{Leconte}, S.~N. {Raymond}, F.~{Selsis}, M.~{Turbet},
  K.~{Barkaoui}, A.~{Burgasser}, M.~R. {Burleigh}, S.~J. {Carey},
  A.~{Chaushev}, C.~M. {Copperwheat}, L.~{Delrez}, C.~S. {Fernandes}, D.~L.
  {Holdsworth}, E.~J. {Kotze}, V.~{Van Grootel}, Y.~{Almleaky},
  Z.~{Benkhaldoun}, P.~{Magain}, D.~{Queloz}, {Seven temperate terrestrial
  planets around the nearby ultracool dwarf star TRAPPIST-1}.
\newblock {\it Nature\/} {\bf 542}, 456-460 (2017).

\bibitem{Dittmann2017}
J.~A. {Dittmann}, J.~M. {Irwin}, D.~{Charbonneau}, X.~{Bonfils},
  N.~{Astudillo-Defru}, R.~D. {Haywood}, Z.~K. {Berta-Thompson}, E.~R.
  {Newton}, J.~E. {Rodriguez}, J.~G. {Winters}, T.-G. {Tan}, J.-M. {Almenara},
  F.~{Bouchy}, X.~{Delfosse}, T.~{Forveille}, C.~{Lovis}, F.~{Murgas},
  F.~{Pepe}, N.~C. {Santos}, S.~{Udry}, A.~{W{\"u}nsche}, G.~A. {Esquerdo},
  D.~W. {Latham}, C.~D. {Dressing}, {A temperate rocky super-Earth transiting a
  nearby cool star}.
\newblock {\it Nature\/} {\bf 544}, 333-336 (2017).

\bibitem{Patel2015}
B.~H. {Patel}, C.~{Percivalle}, D.~J. {Ritson}, C.~D. {Duffy}, J.~D.
  {Sutherland}, {Common origins of RNA, protein and lipid precursors in a
  cyanosulfidic protometabolism}.
\newblock {\it Nature Chemistry\/} {\bf 7}, 301-307 (2015).

\bibitem{Xu2018}
J.~Xu, D.~J. Ritson, S.~Ranjan, Z.~R. Todd, D.~D. Sasselov, J.~D. Sutherland,
  {Photochemical reductive homologation of hydrogen cyanide using sulfite and ferrocyanide}.
\newblock {\it Chemical Communications\/} 10.1039/C8CC01499J  (2018).

\bibitem{Sutherland2017}
J.~D. Sutherland, Opinion: Studies on the origin of life—the end of the
  beginning.
\newblock {\it Nature Reviews Chemistry\/} {\bf 1}, 0012 (2017).

\bibitem{Ranjan2016}
S.~{Ranjan}, D.~D. {Sasselov}, {Influence of the UV Environment on the
  Synthesis of Prebiotic Molecules}.
\newblock {\it Astrobiology\/} {\bf 16}, 68-88 (2016).

\bibitem{Todd2018}
Z.~R. Todd, A.~C. Fahrenbach, C.~J. Magnani, S.~Ranjan, A.~Bj{\"o}rkbom, J.~W.
  Szostak, D.~D. Sasselov, Solvated-electron production using cyanocuprates is
  compatible with the uv-environment on a hadean--archaean earth.
\newblock {\it Chemical Communications\/}  (2018).

\bibitem{Rugheimer2015b}
S.~{Rugheimer}, A.~{Segura}, L.~{Kaltenegger}, D.~{Sasselov}, {UV Surface
  Environment of Earth-like Planets Orbiting FGKM Stars through Geological
  Evolution}.
\newblock {\it ApJ\/} {\bf 806}, 137 (2015).

\bibitem{Ranjan2017c}
S.~{Ranjan}, R.~{Wordsworth}, D.~D. {Sasselov}, {The Surface UV Environment on
  Planets Orbiting M-Dwarfs: Implications for Prebiotic Chemistry and the Need
  for Experimental Follow-up}.
\newblock {\it ApJ\/} {\bf 843}, 110 (2017).

\bibitem{Crispino1993}
G.~A. Crispino, P.~T. Ho, K.~B. Sharpless, Selective perhydroxylation of
  squalene: taming the arithmetic demon.
\newblock {\it Science\/} {\bf 259}, 64--66 (1993).


\bibitem{Kasting1993}
J.~F. {Kasting}, {Earth's early atmosphere}.
\newblock {\it Science\/} {\bf 259}, 920-926 (1993).

\bibitem{Rogers2015}
L.~A.~{Rogers}, {Most 1.6 Earth-radius Planets are Not Rocky}.
\newblock {\it Astrophysical Journal\/} {\bf 801}, 41 (2015).

\bibitem{Wolfgang2015}
A.~{Wolfgang}, E.~{Lopez}, {How Rocky Are They? The Composition Distribution of Kepler's Sub-Neptune Planet Candidates within 0.15 AU}.
\newblock {\it Astrophysical Journal} {\bf 806}, 183 (2015).

\bibitem{Chen2017}
J.~{Chen}, D.~{Kipping}, {Forecasted masses for seven thousand KOIs},
\newblock {\it Monthly Notices of the Royal Astronomical Society} {\bf 473}, 2753 (2017).

\bibitem{Morton2016}
T.~D. {Morton}, S.~T. {Bryson}, J.~L. {Coughlin}, J.~F. {Rowe},
  G.~{Ravichandran}, E.~A. {Petigura}, M.~R. {Haas}, N.~M. {Batalha}, {False
  Positive Probabilities for all Kepler Objects of Interest: 1284 Newly
  Validated Planets and 428 Likely False Positives}.
\newblock {\it ApJ\/} {\bf 822}, 86 (2016).

\bibitem{Angelo2017}
I.~{Angelo}, J.~F. {Rowe}, S.~B. {Howell}, E.~V. {Quintana}, M.~{Still}, A.~W.
  {Mann}, B.~{Burningham}, T.~{Barclay}, D.~R. {Ciardi}, D.~{Huber}, S.~R.
  {Kane}, {Kepler-1649b: An Exo-Venus in the Solar Neighborhood}.
\newblock {\it AJ\/} {\bf 153}, 162 (2017).

\bibitem{Koppa2013}
R.~K. {Kopparapu}, R.~{Ramirez}, J.~F. {Kasting}, V.~{Eymet}, T.~D. {Robinson},
  S.~{Mahadevan}, R.~C. {Terrien}, S.~{Domagal-Goldman}, V.~{Meadows},
  R.~{Deshpande}, {Habitable Zones around Main-sequence Stars: New Estimates}.
\newblock {\it ApJ\/} {\bf 765}, 131 (2013).

\bibitem{Koppa2014}
R.~K. {Kopparapu}, R.~M. {Ramirez}, J.~{SchottelKotte}, J.~F. {Kasting},
  S.~{Domagal-Goldman}, V.~{Eymet}, {Habitable Zones around Main-sequence
  Stars: Dependence on Planetary Mass}.
\newblock {\it ApJ Letters\/} {\bf 787}, L29 (2014).

\bibitem{Davenport16}
J.~R.~A. {Davenport}, {The Kepler Catalog of Stellar Flares}.
\newblock {\it Astrophys. J.\/} {\bf 829}, 23 (2016).

\bibitem{Air2016}
V.~S. Airapetian, A.~Glocer, G.~Gronoff, E.~Hebrard, W.~Danchi, Prebiotic
  chemistry and atmospheric warming of early earth by an active young sun.
\newblock {\it Nature Geosci\/} {\bf 9}, 452--455 (2016).

\bibitem{Xu2016}
J.~Xu, M.~Tsanakopoulou, C.~J. Magnani, R.~Szabla, J.~E. {\v{S}}poner,
  J.~{\v{S}}poner, R.~W. G{\'o}ra, J.~D. Sutherland, A prebiotically plausible
  synthesis of pyrimidine $\beta$-ribonucleosides and their phosphate
  derivatives involving photoanomerization.
\newblock {\it Nature Chemistry\/}  (2016).

\bibitem{Forgan2013}
D.~{Forgan}, D.~{Kipping}, {Dynamical effects on the habitable zone for
  Earth-like exomoons}.
\newblock {\it MNRAS\/} {\bf 432}, 2994-3004 (2013).

\bibitem{Cleaves2008}
H.~J. {Cleaves}, J.~H. {Chalmers}, A.~{Lazcano}, S.~L. {Miller}, J.~L. {Bada},
  {A Reassessment of Prebiotic Organic Synthesis in Neutral Planetary
  Atmospheres}.
\newblock {\it Origins of Life and Evolution of the Biosphere\/} {\bf 38},
  105-115 (2008).

\bibitem{Ruiz2014}
K.~Ruiz-Mirazo, C.~Briones, A.~de~la Escosura, Prebiotic systems chemistry: new
  perspectives for the origins of life.
\newblock {\it Chem. Rev\/} {\bf 114}, 285--366 (2014).

\bibitem{Bernstein2002}
M.~P. {Bernstein}, J.~P. {Dworkin}, S.~A. {Sandford}, G.~W. {Cooper}, L.~J.
  {Allamandola}, {Racemic amino acids from the ultraviolet photolysis of
  interstellar ice analogues}.
\newblock {\it Nature\/} {\bf 416}, 401-403 (2002).

\bibitem{Rushby2013}
A.~J. {Rushby}, M.~W. {Claire}, H.~{Osborn}, A.~J. {Watson}, {Habitable Zone
  Lifetimes of Exoplanets around Main Sequence Stars}.
\newblock {\it Astrobiology\/} {\bf 13}, 833-849 (2013).

\bibitem{France2016}
K.~{France}, R.~O. {Parke Loyd}, A.~{Youngblood}, A.~{Brown}, P.~C.
  {Schneider}, S.~L. {Hawley}, C.~S. {Froning}, J.~L. {Linsky}, A.~{Roberge},
  A.~P. {Buccino}, J.~R.~A. {Davenport}, J.~M. {Fontenla}, L.~{Kaltenegger},
  A.~F. {Kowalski}, P.~J.~D. {Mauas}, Y.~{Miguel}, S.~{Redfield},
  S.~{Rugheimer}, F.~{Tian}, M.~C. {Vieytes}, L.~M. {Walkowicz}, K.~L.
  {Weisenburger}, {The MUSCLES Treasury Survey. I. Motivation and Overview}.
\newblock {\it ApJ\/} {\bf 820}, 89 (2016).

\bibitem{Youngblood2016}
A.~{Youngblood}, K.~{France}, R.~O. {Parke Loyd}, J.~L. {Linsky},
  S.~{Redfield}, P.~C. {Schneider}, B.~E. {Wood}, A.~{Brown}, C.~{Froning},
  Y.~{Miguel}, S.~{Rugheimer}, L.~{Walkowicz}, {The MUSCLES Treasury Survey.
  II. Intrinsic LY{$\alpha$} and Extreme Ultraviolet Spectra of K and M Dwarfs
  with Exoplanets*}.
\newblock {\it ApJ\/} {\bf 824}, 101 (2016).

\bibitem{Loyd2016}
R.~O.~P. {Loyd}, K.~{France}, A.~{Youngblood}, C.~{Schneider}, A.~{Brown},
  R.~{Hu}, J.~{Linsky}, C.~S. {Froning}, S.~{Redfield}, S.~{Rugheimer},
  F.~{Tian}, {The MUSCLES Treasury Survey. III. X-Ray to Infrared Spectra of 11
  M and K Stars Hosting Planets}.
\newblock {\it ApJ\/} {\bf 824}, 102 (2016).

\bibitem{Ribas2010}
I.~{Ribas}, G.~F. {Porto de Mello}, L.~D. {Ferreira}, E.~{H{\'e}brard},
  F.~{Selsis}, S.~{Catal{\'a}n}, A.~{Garc{\'e}s}, J.~D. {do Nascimento}, Jr.,
  J.~R. {de Medeiros}, {Evolution of the Solar Activity Over Time and Effects
  on Planetary Atmospheres. II. {$\kappa$}$^{1}$ Ceti, an Analog of the Sun
  when Life Arose on Earth}.
\newblock {\it ApJ\/} {\bf 714}, 384-395 (2010).

\bibitem{Segura2005}
A.~{Segura}, J.~F. {Kasting}, V.~{Meadows}, M.~{Cohen}, J.~{Scalo}, D.~{Crisp},
  R.~A.~H. {Butler}, G.~{Tinetti}, {Biosignatures from Earth-Like Planets
  Around M Dwarfs}.
\newblock {\it Astrobiology\/} {\bf 5}, 706-725 (2005).

\bibitem{Rugheimer2013}
S.~{Rugheimer}, L.~{Kaltenegger}, A.~{Zsom}, A.~{Segura}, D.~{Sasselov},
  {Spectral Fingerprints of Earth-like Planets Around FGK Stars}.
\newblock {\it Astrobiology\/} {\bf 13}, 251-269 (2013).

\bibitem{Rimmer2016}
P.~B. {Rimmer}, C.~{Helling}, {A Chemical Kinetics Network for Lightning and
  Life in Planetary Atmospheres}.
\newblock {\it ApJ Supplement Series\/} {\bf 224}, 9 (2016).

\bibitem{Gaillard2014}
F.~{Gaillard}, B.~{Scaillet}, {A theoretical framework for volcanic degassing
  chemistry in a comparative planetology perspective and implications for
  planetary atmospheres}.
\newblock {\it Earth and Planetary Science Letters\/} {\bf 403}, 307-316
  (2014).

\bibitem{Hu2014}
R.~{Hu}, S.~{Seager}, {Photochemistry in Terrestrial Exoplanet Atmospheres.
  III. Photochemistry and Thermochemistry in Thick Atmospheres on Super Earths
  and Mini Neptunes}.
\newblock {\it Astrophysical Journal\/} {\bf 784}, 63 (2014).

\bibitem{Toon1989}
O.~B. {Toon}, C.~P. {McKay}, T.~P. {Ackerman}, K.~{Santhanam}, {Rapid
  calculation of radiative heating rates and photodissociation rates in
  inhomogeneous multiple scattering atmospheres}.
\newblock {\it Journal of Geophysical Research: Atmospheres\/} {\bf 94},
  16287-16301 (1989).

\bibitem{Sauer2004}
M.~C. {Sauer}, R.~A. {Crowell}, I.~A. {Shkrob}, {Electron Photodetachment from
  Aqueous Anions. 1. Quantum Yields for Generation of Hydrated Electron by 193
  and 248 nm Laser Photoexcitation of Miscellaneous Inorganic Anions}.
\newblock {\it Journal of Physical Chemistry A\/} {\bf 108}, 5490-5502 (2004).

\bibitem{Hawley1991}
S.~L. {Hawley}, B.~R. {Pettersen}, {The great flare of 1985 April 12 on AD
  Leonis}.
\newblock {\it Astrophys. J.\/} {\bf 378}, 725-741 (1991).

\bibitem{Segura2010}
A.~{Segura}, L.~M. {Walkowicz}, V.~{Meadows}, J.~{Kasting}, S.~{Hawley}, {The
  Effect of a Strong Stellar Flare on the Atmospheric Chemistry of an
  Earth-like Planet Orbiting an M Dwarf}.
\newblock {\it Astrobiology\/} {\bf 10}, 751-771 (2010).

\bibitem{Hawley14}
S.~L. {Hawley}, J.~R.~A. {Davenport}, A.~F. {Kowalski}, J.~P. {Wisniewski},
  L.~{Hebb}, R.~{Deitrick}, E.~J. {Hilton}, {Kepler Flares. I. Active and
  Inactive M Dwarfs}.
\newblock {\it Astrophys. J.\/} {\bf 797}, 121 (2014).

\bibitem{Hilton2011}
E.~J. {Hilton}, {The Galactic M Dwarf Flare Rate}, Ph.D. thesis, University of
  Washington (2011).

\bibitem{Davenport14}
J.~R.~A. {Davenport}, S.~L. {Hawley}, L.~{Hebb}, J.~P. {Wisniewski}, A.~F.
  {Kowalski}, E.~C. {Johnson}, M.~{Malatesta}, J.~{Peraza}, M.~{Keil}, S.~M.
  {Silverberg}, T.~C. {Jansen}, M.~S. {Scheffler}, J.~R. {Berdis}, D.~M.
  {Larsen}, E.~J. {Hilton}, {Kepler Flares. II. The Temporal Morphology of
  White-light Flares on GJ 1243}.
\newblock {\it Astrophys. J.\/} {\bf 797}, 122 (2014).

\bibitem{Kirk16}
B.~{Kirk}, K.~{Conroy}, A.~{Pr{\v s}a}, M.~{Abdul-Masih}, A.~{Kochoska},
  G.~{Matijevi{\v c}}, K.~{Hambleton}, T.~{Barclay}, S.~{Bloemen},
  T.~{Boyajian}, L.~R. {Doyle}, B.~J. {Fulton}, A.~J. {Hoekstra}, K.~{Jek},
  S.~R. {Kane}, V.~{Kostov}, D.~{Latham}, T.~{Mazeh}, J.~A. {Orosz},
  J.~{Pepper}, B.~{Quarles}, D.~{Ragozzine}, A.~{Shporer}, J.~{Southworth},
  K.~{Stassun}, S.~E. {Thompson}, W.~F. {Welsh}, E.~{Agol}, A.~{Derekas},
  J.~{Devor}, D.~{Fischer}, G.~{Green}, J.~{Gropp}, T.~{Jacobs}, C.~{Johnston},
  D.~M. {LaCourse}, K.~{Saetre}, H.~{Schwengeler}, J.~{Toczyski}, G.~{Werner},
  M.~{Garrett}, J.~{Gore}, A.~O. {Martinez}, I.~{Spitzer}, J.~{Stevick}, P.~C.
  {Thomadis}, E.~H. {Vrijmoet}, M.~{Yenawine}, N.~{Batalha}, W.~{Borucki},
  {Kepler Eclipsing Binary Stars. VII. The Catalog of Eclipsing Binaries Found
  in the Entire Kepler Data Set}.
\newblock {\it AJ\/} {\bf 151}, 68 (2016).
\end{thebibliography}

\begin{scilastnote}
\item {\bf Funding:} All authors thank the Simons Foundation and Kavli Foundation for funding. {\bf Other Acknowledgements:} We thank the three referees for their helpful comments. P.B.R. thanks Kevin France and Vladimir Airipitian for help with stellar UV and particle spectra, and P.B.R. and J.S. thank Dougal Ritson for key insights regarding the hydrogen sulfide chemistry. {\bf Author Contributions:} P.B.R. came up with the original idea for the paper, wrote the majority of the paper and portions of the Methods and performed the analysis of the chemical kinetics and surface UV fluxes. J.X. and P.B.R. performed all the laboratory experiments together and J.X. performed the NMR analysis and wrote the Supplementary Information. S.J.T. and P.B.R. measured the UV output of the mercury lamps together, and S.J.T. wrote the Methods section regarding the UV lamps. E.G. performed the flare analysis and wrote the relevant section in the main text and the Methods section. J.D.S. identified the key light and dark reactions to consider, D.Q. identified appropriate criteria for the planets we consider and helped design Fig. 4. Both J.D.S. and D.Q. participated in writing portions of the main text, and edited the entire paper for accuracy and clarity. {\bf Competing interests:} The authors declare that they have no competing interests. {\bf Data and materials availability:} Additional data related to this paper may be requested from the authors.
\end{scilastnote}

\end{document}


\begin{efigure}[h!]
\begin{center}
\includegraphics[width=9cm]{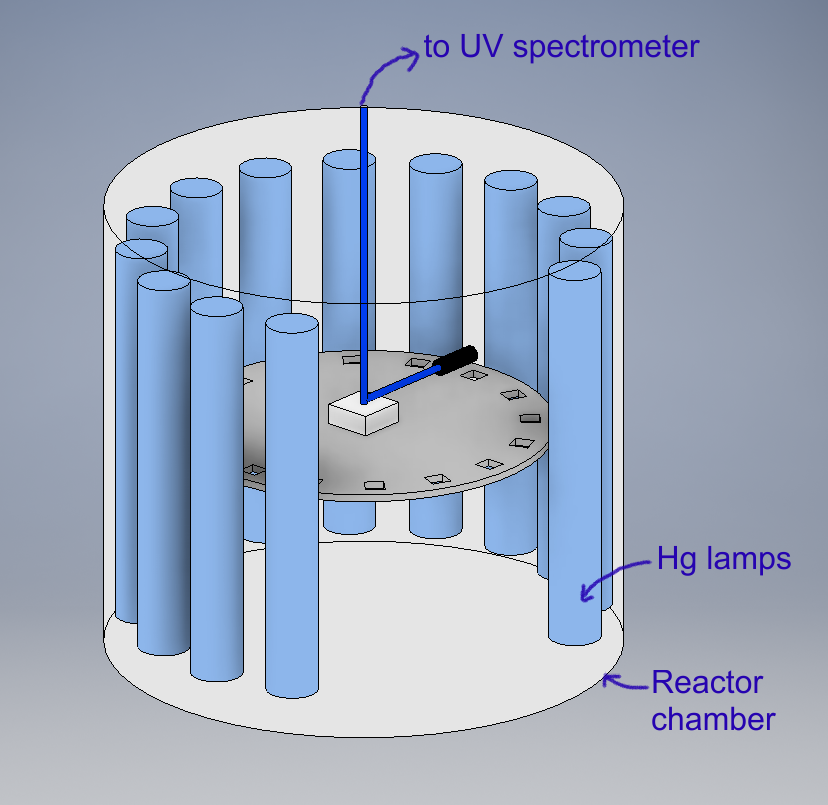}
\caption{{\bf The UV Reactor.} Schematic of the UV reactor chamber containing 16 Hg lamps (3 not shown for clarity).  An optical fibre connected to the UV spectrometer is placed down the centre axis of the chamber with the probe end of the fibre located at the approximate centre of the quartz cuvette.}
\label{fig:uv_chamber}
\end{center}
\end{efigure}

\begin{efigure}[h!]
\begin{center}
\def\big{\includegraphics[height=9cm]{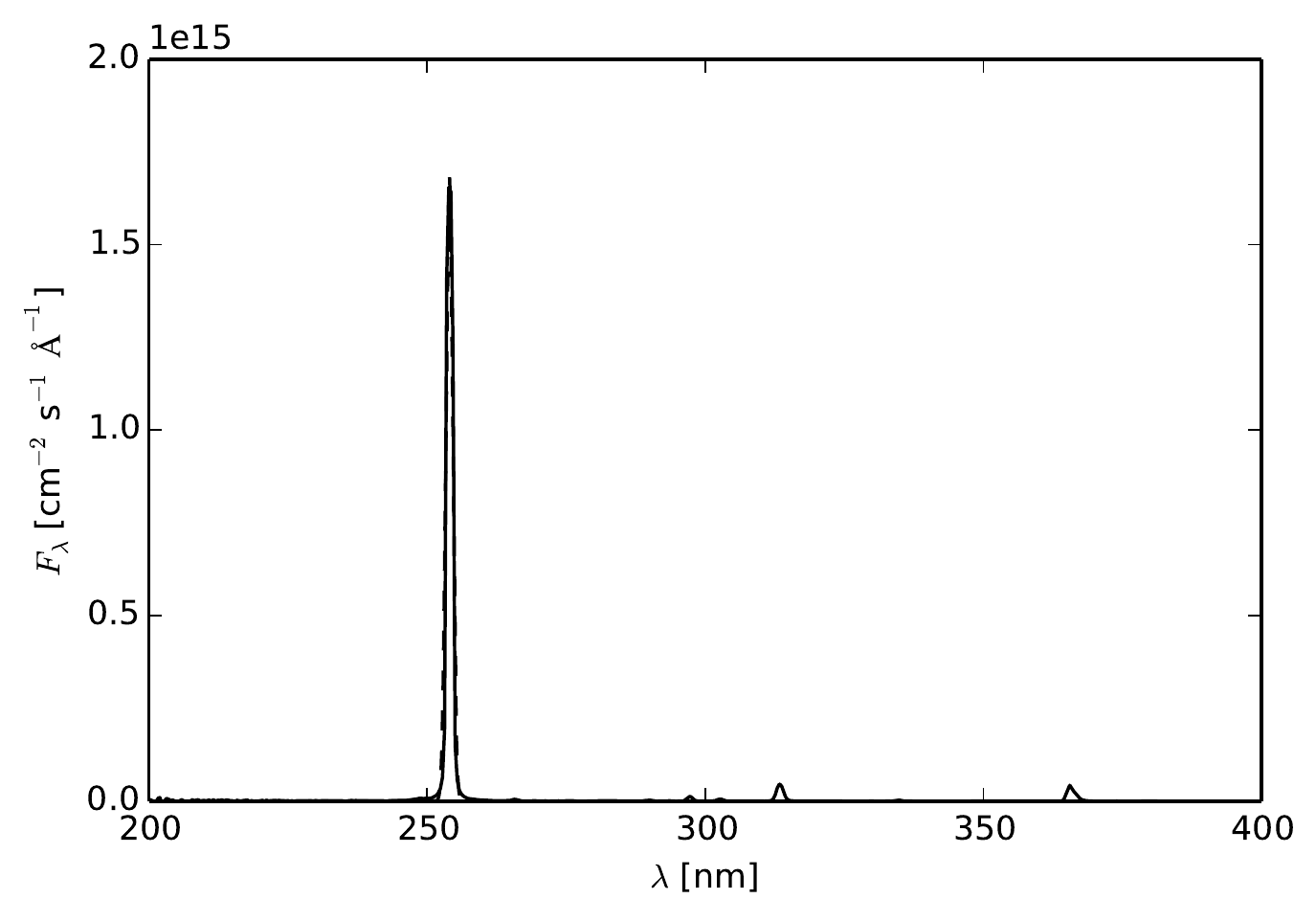}}
\def\little{\includegraphics[height=6cm]{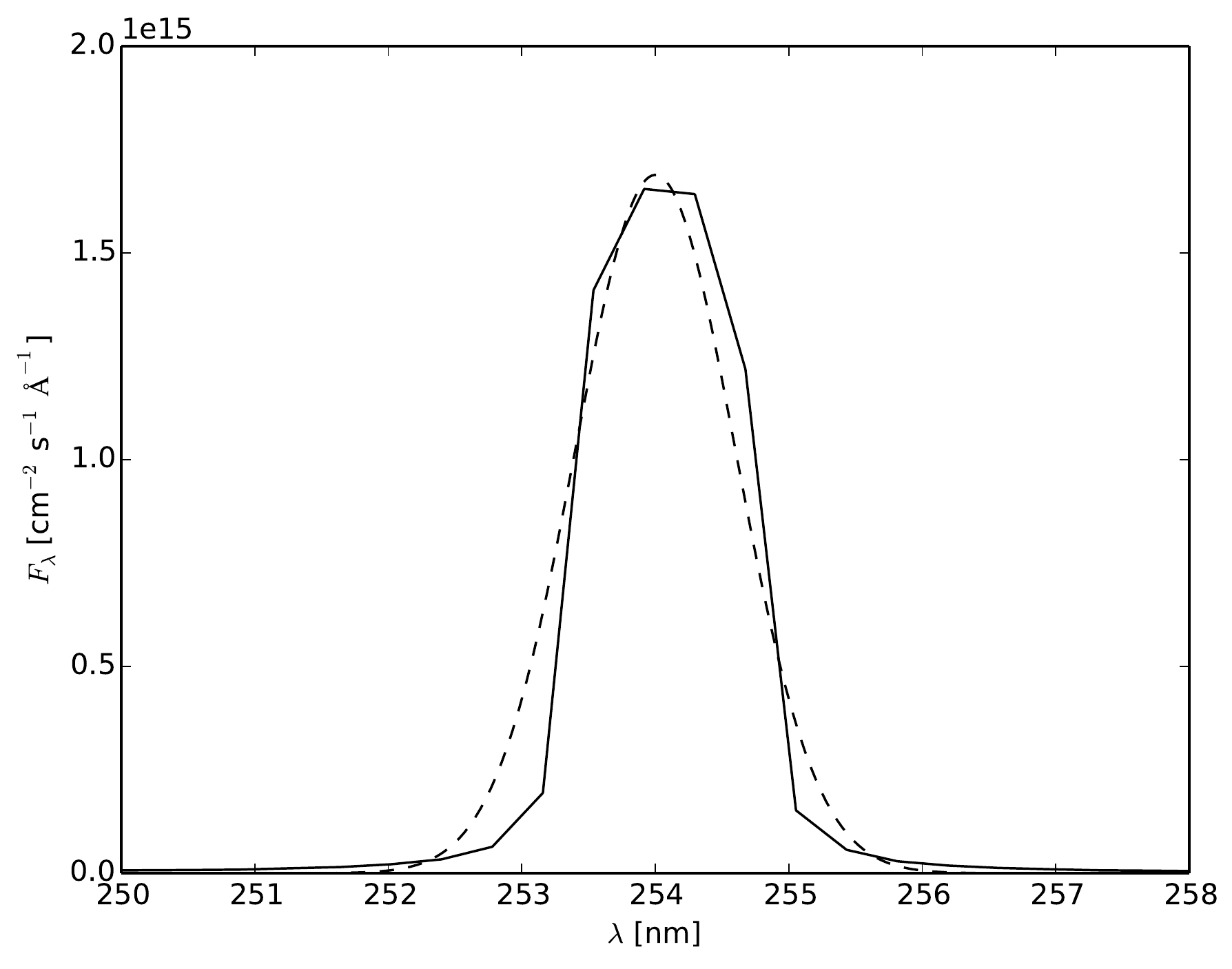}}
\def\stackalignment{r}
\topinset{\little}{\big}{-7pt}{-40pt}
\caption{{\bf Reactor Emission} Plot of the measured (solid line) emission from the Hg lamps over the 200-400 nm spectral range.  The inset plot is a detailed view around the principle emission line at 254 nm which contains $\sim 90$\% of the total flux.  The dashed line shows the fitted profile for the 254 nm emission line.}
\label{fig:spec_UVlamp}
\end{center}
\end{efigure}

\begin{efigure}[h!]
\includegraphics[width=\textwidth]{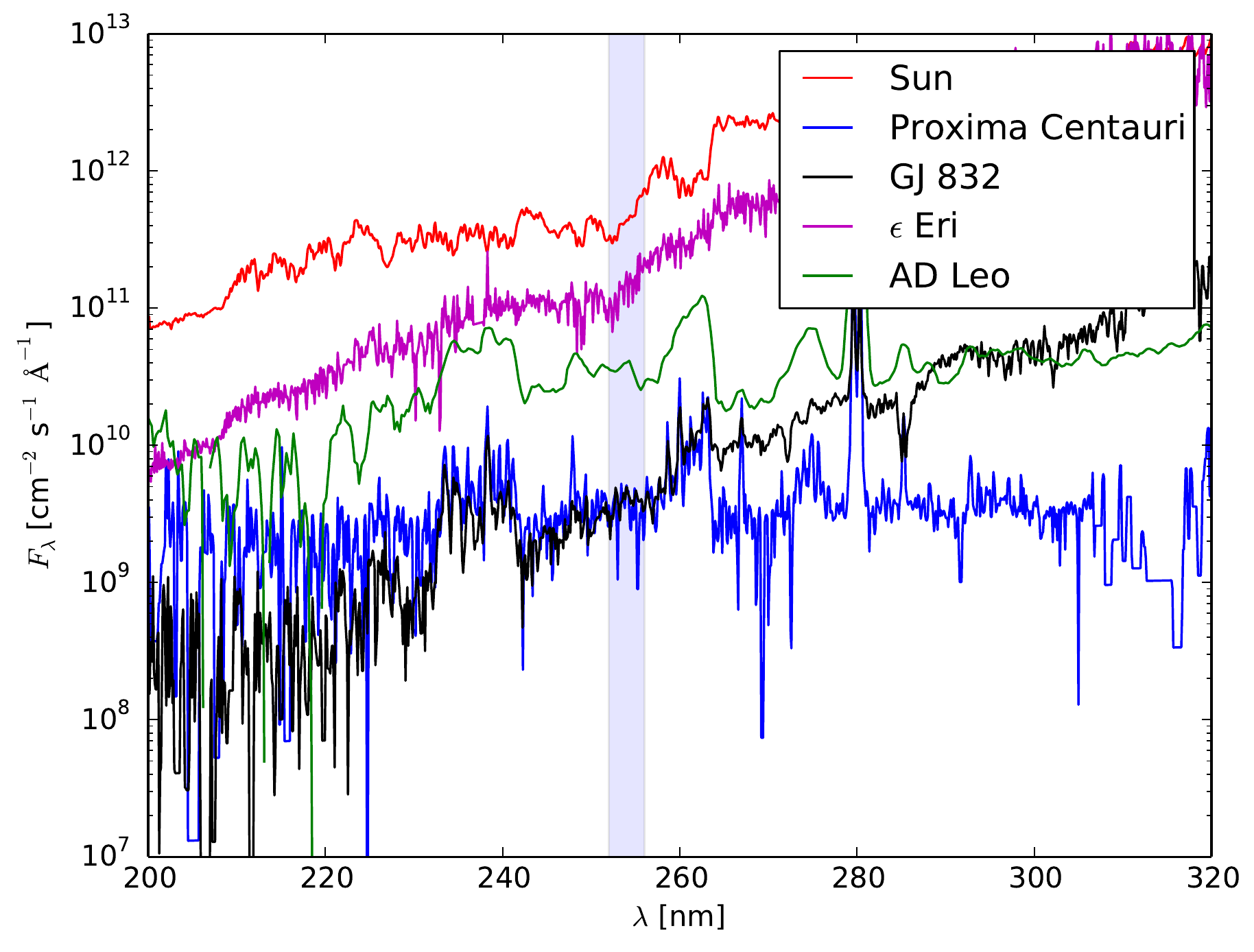}
\caption{{\bf Stellar Spectra.} The UV spectral irradiance as a function of wavelength for the Young Sun (32), AD Leo (33), and MUSCLES spectra of Proxima Centauri, GJ 832 and $\epsilon$ Eri (29-31). These spectra are for a hypothetical planet within the habitable zone of these stars, with no atmospheric extinction. These fluxes are used for the UV phototransport equation as the boundary condition for the top of the atmosphere.}
\label{fig:UVcomp}
\end{efigure}

\begin{efigure}[h!]
\includegraphics[width=\textwidth]{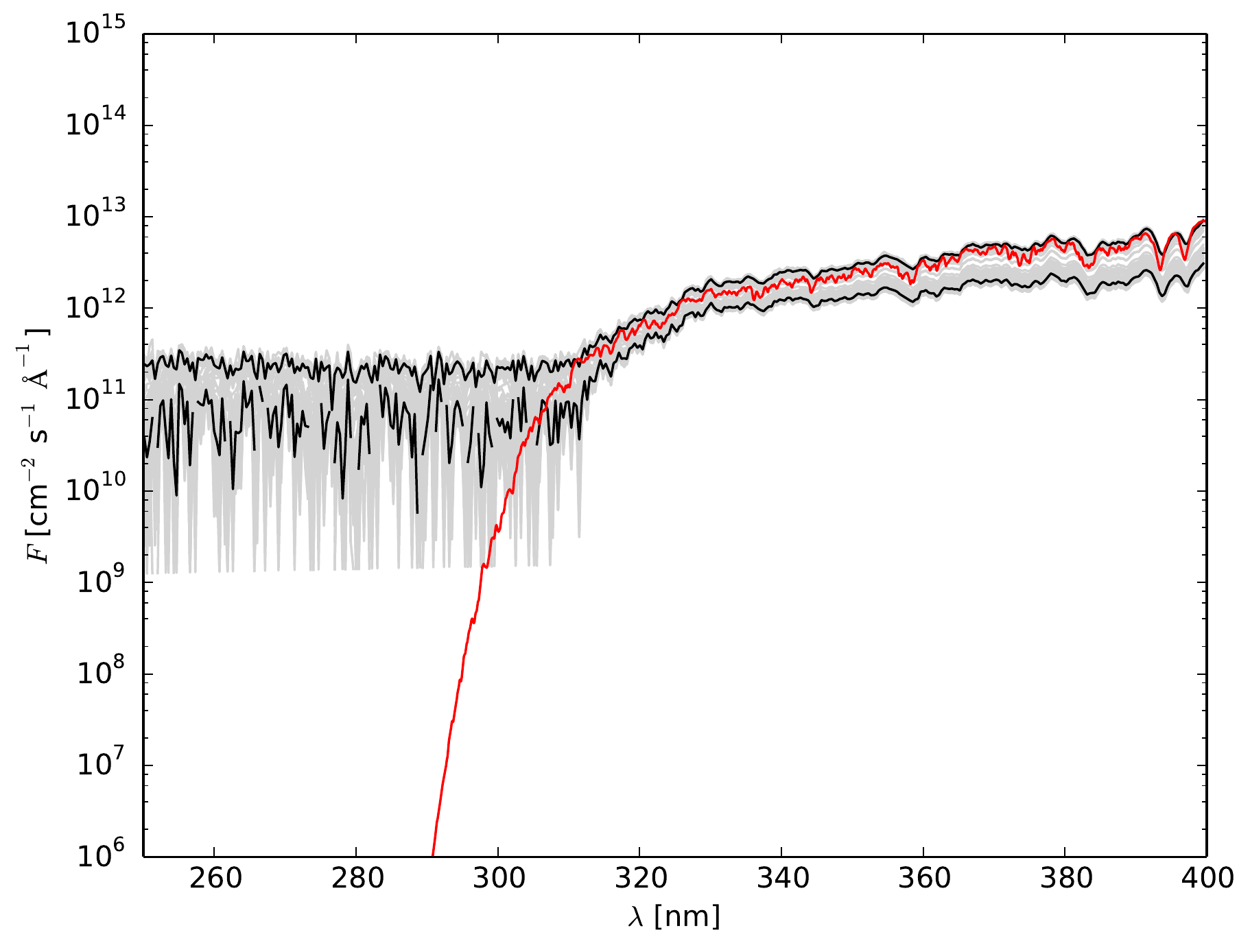}
\caption{{\bf Surface spectral irradiance at Cambridge.} A comparison between the ARGO model surface spectral irradiance as a function of wavelength at Cambridge on 24 February 2017 at 15:00, compared to observations on this date.}
\label{fig:earthmodel}
\end{efigure}

\begin{efigure}[h!]
\includegraphics[width=\textwidth]{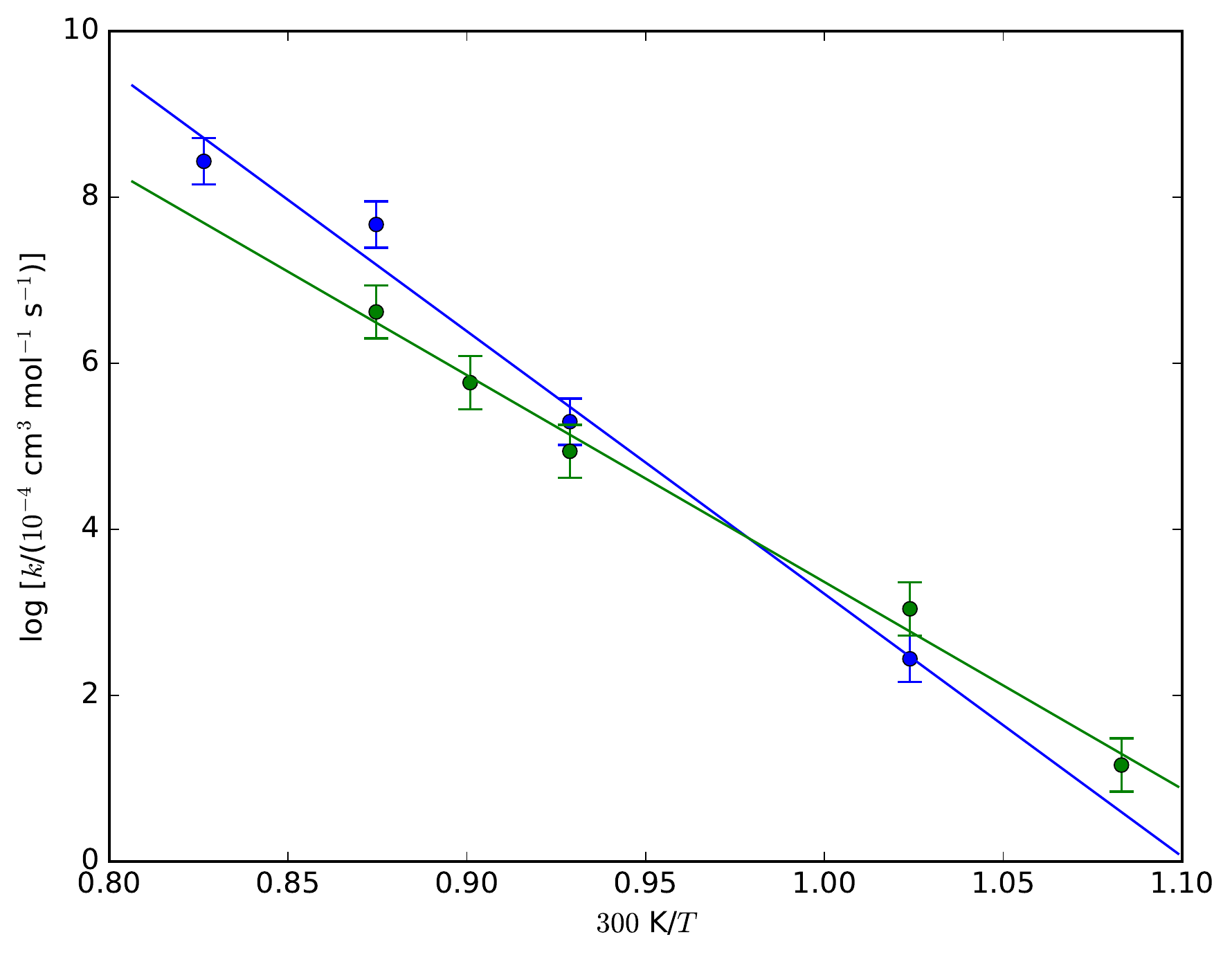}
\caption{{\bf Ahrennius Plots.} The Ahrennius Plot of the log of the rate constant as a function of 300 K/$T$, for bisulfite (Eq. (14) and (15), blue line) and hydrogen sulfide (Eq. (16) and (17)). \label{fig:Arrhenius}}
\end{efigure}

\begin{efigure}
\includegraphics[width=\linewidth]{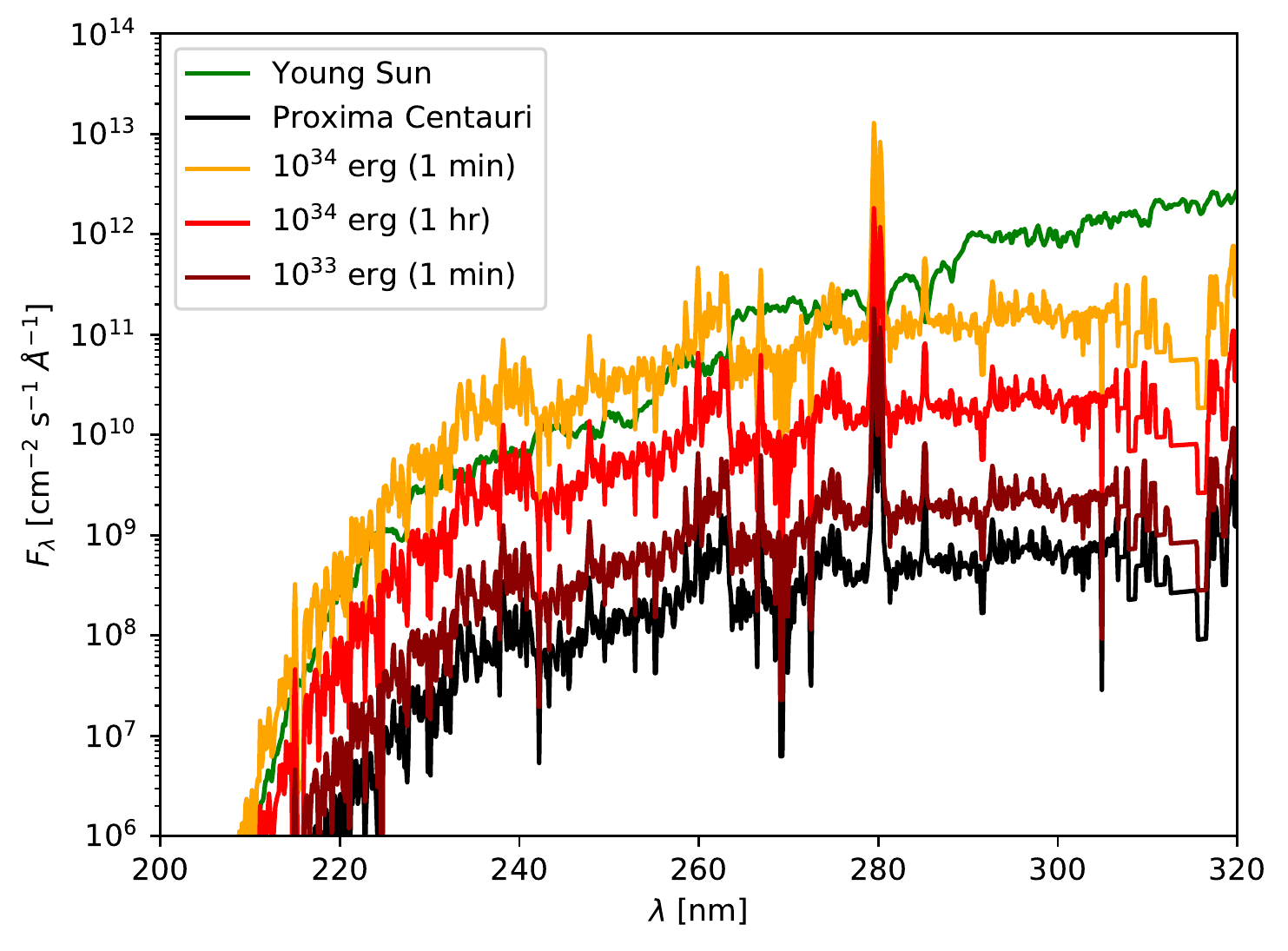}
\caption{{\bf Flare Spectra.} Same as Fig. 2 for Proxima b, with the young sun as a point of comparison. The plot shows the effect a flare of energy $10^{33}$ erg and $10^{34}$ erg has on the flux after 1 minute and 1 hour; the $10^{33}$ erg spectrum after 1 hour is virtually indistinguishable from the quiescent stellar flux on a log scale.}
\label{fig:UV-flare}
\end{efigure}

\clearpage

\begin{etable}[h!]
\begin{center}
\caption{HCN + SO$_3^{2-}$ \label{tab:bisulfite}}
 \begin{tabular}{c c c c} 
 \hline
 $T$ [K] & $t$ [h] & $c$ [mM] & k [mol$^{-1}$ L s$^{-1}$] \\ [0.5ex] 
 \hline\hline
 293 & 48 & 8 & $1.2 \times 10^{-6}$ \\ 
 \hline
 293 & 96 & 14 & $1.1 \times 10^{-6}$  \\
 \hline
 323 & 1.5 & 4 & $2.0 \times 10^{-5}$  \\
 \hline
 343 & 1 & 30 & $2.4 \times 10^{-4}$ \\
 \hline
 343 & 2 & 40 & $1.9 \times 10^{-4}$ \\
 \hline
 363 & 1.5 & 90 & $4.6 \times 10^{-4}$ \\
 \hline
\end{tabular}
\end{center}
\end{etable}

\begin{etable}[h!]
\begin{center}
\caption{Glycolonitrile + HS$^-$ \label{tab:sulfide}}
 \begin{tabular}{c c c c} 
 \hline
 $T$ [K] & $t$ [h] & $c$ [mM] & k [mol$^{-1}$ L s$^{-1}$] \\ [0.5ex] 
 \hline\hline
 277 & 312 & 13 & $3.2 \times 10^{-7}$ \\ 
 \hline
 293 & 24 & 6 & $2.1 \times 10^{-6}$  \\
 \hline
 323 & 1.5 & 6 & $1.4 \times 10^{-5}$  \\
 \hline
 333 & 2 & 14 & $3.2 \times 10^{-5}$ \\
 \hline
 343 & 1 & 10 & $7.5 \times 10^{-5}$ \\
 \hline
\end{tabular}
\end{center}
\end{etable}

\begin{etable}[h!]
\begin{center}
\caption{Percentages of stars in different spectral type ranges that meet the criterion for activity given by Eq. (38) (defined here as "Active" stars).
\label{tab:lifestars}}
 \begin{tabular}{ccrrrr}
 \hline \hline
 Spectral types  &  Mass range  &  \multicolumn{3}{c}{No. of stars}  &  \% Active\\
   &  & Total  &  Used  &  Active &  \\
 \hline
K5--G8 &  0.7--0.8  &  1215  &  1213  &  231  &  19  \\
K8--K5 &  0.6--0.7  &  117  &  111  &  54  &  49  \\
M1--K8 &  0.5--0.6  &  65  &  56  &  14  &  25  \\
M4--M1 &  0.4--0.5  &  42  &  41  &  4  &  10  \\
$<$M4  &  0.0--0.4  &  6  &  5  &  0  &  0  \\

\hline
\end{tabular}
\end{center}
\end{etable}

\includepdf[pages=-,pagecommand={}]{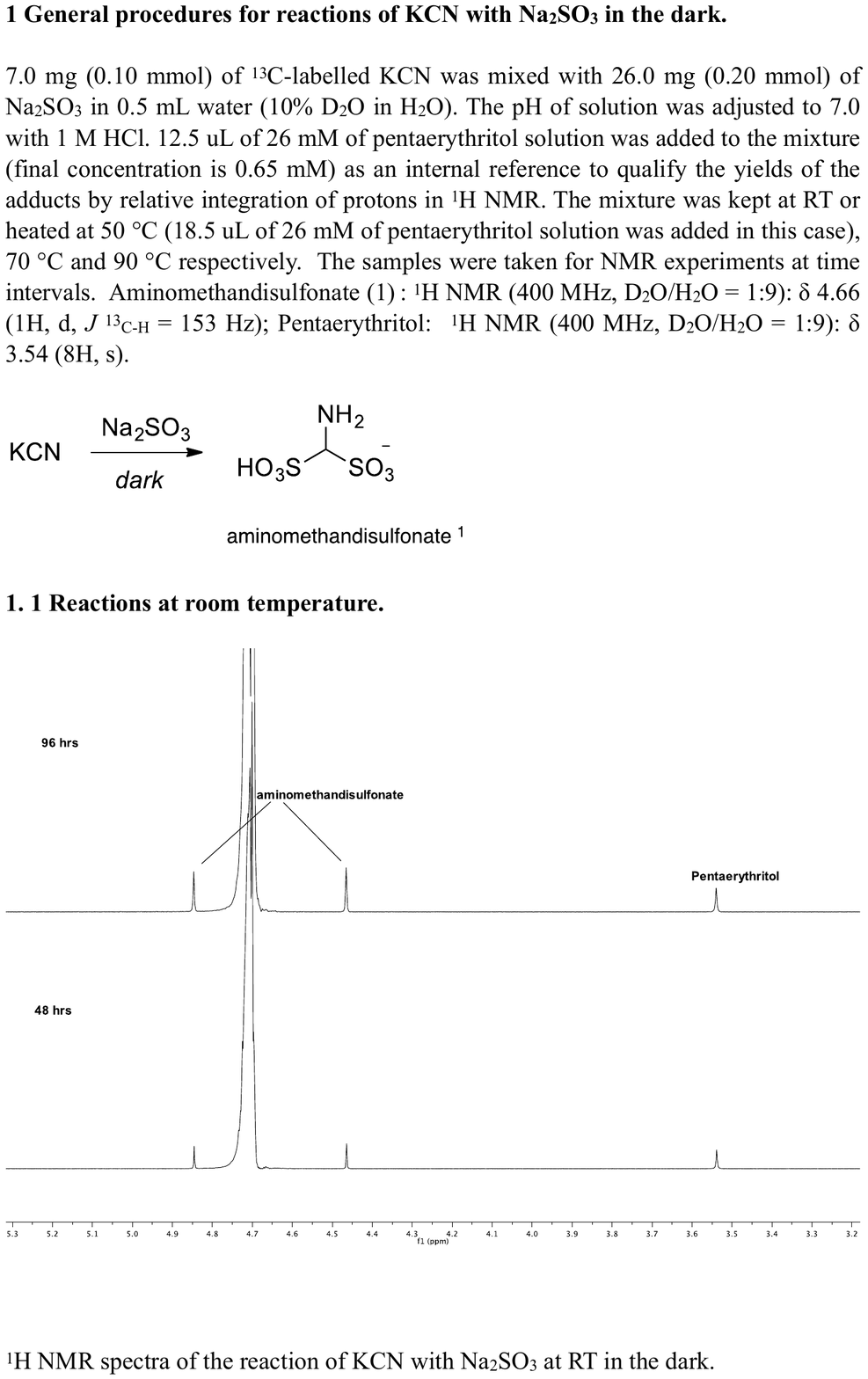}